\documentclass{article}

\usepackage{PRIMEarxiv}

\usepackage{cite}
\usepackage{amsmath,amssymb,amsfonts}
\usepackage{graphicx}
\usepackage{textcomp}

\usepackage{lineno,hyperref}
\usepackage{adjustbox}
\usepackage{multirow}
\usepackage{tablefootnote}
\usepackage{enumitem}
\modulolinenumbers[5]

\usepackage{xcolor,colortbl}
\usepackage{algcompatible}
\usepackage{algorithm}
\usepackage[noend]{algpseudocode}
\mathchardef\mhyphen="2D
\algnewcommand\algorithmicforeach{\textbf{for each}}
\algdef{S}[FOR]{ForEach}[1]{\algorithmicforeach\ #1\ \algorithmicdo}
\algdef{SE}[DOWHILE]{Do}{doWhile}{\algorithmicdo}[1]{\algorithmicwhile\ #1}%
\usepackage{amssymb,amsmath}

\usepackage{url}
\usepackage{hyperref}
\usepackage[none]{hyphenat}

\title{Novel Strategy Generating Variable-length State Machine Test Paths
\thanks{ 
\textbf{Paper accepted for publication in International Journal of Software Engineering and Knowledge Engineering}} 
}

\author{
  Vaclav Rechtberger \\
Dept. of Computer Science, Faculty of Electrical Engineering,\\ Czech Technical University in Prague, Karlovo namesti 13, 121 35, Prague, Czechia\\
\texttt{rechtva1@fel.cvut.cz}
   \And
   Miroslav Bures \\
  Dept. of Computer Science, Faculty of Electrical Engineering,\\ Czech Technical University in Prague, Karlovo namesti 13, 121 35, Prague, Czechia\\
\texttt{miroslav.bures@fel.cvut.cz}\\
   \And
  Bestoun S. Ahmed \\
  Dept of Mathematics and Computer Science, Karlstad University, 651 88 Karlstad, Sweden\\
  Dept of Computer Science, FEE, Czech Technical University in Prague, Czechia \\
  \texttt{bestoun@kau.se} \\
  \And
  Hynek Schvach \\
  Department of Military Medical Service Organisation and Management\\ University of Defence, Trebesska 1575, 500 01, Hradec Kralove, Czechia\\
  \texttt{hynek.schvach@unob.cz} \\
   \And
}

\begin{document}
\maketitle

\begin{abstract}
Finite State Machine is a popular modeling notation for various systems, especially software and electronic. Test paths can be automatically generated from the system model to test such systems using a suitable algorithm. This paper presents a strategy that generates test paths and allows to start and end test paths only in defined states of the finite state machine. The strategy also simultaneously supports generating test paths only of length in a given range. For this purpose, alternative system models, test coverage criteria, and a set of algorithms are developed. The strategy is compared with the best alternative based on the reduction of the test set generated by the established N-switch coverage approach on a mix of 171 industrial and artificially generated problem instances. The proposed strategy outperforms the compared variant in a smaller number of test path steps. The extent varies with the used test coverage criterion and preferred test path length range from none to two and half fold difference. Moreover, the proposed technique detected up to 30\% more simple artificial defects inserted into experimental SUT models per one test step than the compared alternative technique. The proposed strategy is well applicable in situations where a possible test path starts and ends in a state machine needs to be reflected and, concurrently, the length of the test paths has to be in a defined range.
\end{abstract}

% keywords can be removed

\section{Introduction}\label{sec:introduction}

In system testing, test scenarios are usually defined to describe tests performed on a System Under Test (SUT). The core part of a test scenario is a sequence of actions that must be performed during the test \cite{ammann2016introduction}. Such sequences can be created manually by a test analyst based on design documentation of the SUT. They can also be created based on the knowledge of the system's functionality. Such an approach might be ineffective for complex systems and is prone to design defects in test scenarios. The approach may also lead to a set of test scenarios, for which it is not clear the tests cover which parts of the SUT. The test design process can be automated to minimize these drawbacks, subject to the established Model-Based Testing (MBT) discipline \cite{utting2012taxonomy,ahmad2019model}.

%%%%%%%%%%% Bestoun --- 25 November

In MBT, part or viewpoint on the SUT is modeled by a suitable notation. Then, given the required test coverage criteria, test paths are generated from the model using dedicated algorithms \cite{li2017survey}. A variety of SUT modeling notations can be used within MBT, e.g., \cite{ahmad2019model,li2017survey}. This study focuses on Finite State Machine (FSM) to model parts of the SUT. FSM and its variants are widely used in industry as a modeling notation for system modeling and testing. FSM describes a natural aspect of a wide variety of systems, modules, or data objects processed by a system that is switching from state to state. Therefore, FSMs are one of the fundamental modeling options. Here, FSM-based testing is one of the key testing methods \cite{simao2009comparing}. 

%%%%%%%%%%% Bestoun --- 25 November

In real-world projects, budgets and testing time are typically limited \cite{jones2011economics}. Certain pragmatism and prioritization in covering essential parts of the SUT by more thorough tests and less critical ones by more lightweight tests are highly desirable here. From the test practitioner's viewpoint, it is the foremost approach to achieve the best quality of a SUT given these budget and time constraints. Test coverage criteria are used to determine the thoroughness of the paths \cite{ammann2016introduction}. However, determining the suitable coverage criteria is tricky here. In addition, a careful algorithmic design and implementation are needed to generate test paths that meet the designed coverage criteria. To this end, several test coverage criteria and algorithms have been defined and examined for FSMs in the literature. A careful analysis of the literature is given in Section \ref{sec:background}. 

With the availability of several strategies and studies, the current techniques do not sufficiently reflect the fact that the test paths can effectively start and end only in certain states of the FSM. Attempting to start or end a test in certain states might be highly ineffective or even futile \cite{derderian2010estimating,kalaji2009generating}. In addition, the test path length is less explored in the literature, which is more critical from the test practitioner's perspective. Too long test paths are hard to maintain, and if interrupted by a defect in an SUT, it is tricky to test the rest of the flow \footnote{https://dzone.com/articles/17-best-tips-to-write-effective-test-cases or https://reqtest.com/testing-blog/learn-how-to-write-effective-test-cases/ to give few examples}. Also, too short test paths might be ineffective because of related overhead implied by the execution of the test paths. For example, putting the SUT into an initial state, cleaning procedures after tests, or test reporting effort \cite{vroon2013tmap}.

Considering the aforementioned gaps in the literature in a new strategy will lead to an effort-effective method of FSM testing. To this end, this paper presents a strategy that supports both requirements and compares it with an ad-hoc approach based on an established N-switch test path concept. The contributions of the paper are as follows: 
(1) Novel strategy that generates FSM test paths concurrently, allowing one to limit their length and express in which states of FSM a test path can start and end is presented, (2) a comparable strategy based on the established N-switch concept is presented, and (3) both strategies are compared using several criteria, including their effectiveness in detecting artificial defects in SUT models, and the results are discussed.

This paper is organized as follows. Section \ref{sec:background} introduces the formal preliminaries used in this work and summarizes related works. Section \ref{sec:proposed_approach} presents the proposed FSM test path generation strategy, starting with an SUT model and the definition of the test coverage criteria. The algorithms that generate the test paths are presented here, and our proposal can be compared with an alternative test path generation strategy. Section \ref{sec:initial_experiments} presents the method used in the experiments and their results. Section \ref{sec:threats_to_validity} discussed possible threats to validity and the last section concludes the paper.

\section{Background and Related Work}\label{sec:background}

One of the common notations of FSM used in MBT is based on a directed graph. The typical model is defined as a directed graph $\mathcal{G} = (V,E,v_s,V_e)$ such that $V \neq \emptyset$ is a finite set of vertices representing FSM states and $E \subseteq N \times N$ is a nonempty set of edges $e \in E$ representing FSM transitions. Furthermore, $v_s \in V$ is the start state of the state machine, $V_e \subset V$ is a set of end states of the state machine \cite{wang2019formal,hopcroft2001introduction}. Within this graph, a test path $t$ is a path in $\mathcal{G}$. 

%As commonly accepted, test path does not necessarily have to start in $v_s$ and end in an end vertex from $V_e$. 

Alternatively, part or aspect of the SUT expressed by FSM can be modeled by a Regular Expression (RE) \cite{IntroductionToAutomataTheoryLanguagesAndComputation}. RE describes FSM so that every possible word (sequence of transitions in the FSM) that fits a pattern defined by RE corresponds to a path in this FSM. We denote RE, defining the model as $\boldsymbol{\phi}$. A test path $t$ is a word allowed by $\boldsymbol{\phi}$. In both models, $T$ denotes a set of test paths.

$T$ satisfies \textit{Node Coverage}, when each $v \in V \in \mathcal{G}$ is presented in at least one $t \in T$. In the literature, the Node Coverage is also alternatively denoted as \textit{All States Coverage}. \cite{ammann2016introduction,ModelBasedTestCasesGenerationForOnboardSystem}. %Considering the number of $T$ steps, this criterion is the weakest of all criteria discussed in this study \cite{CoverageCriteriaForStateTransitionTestingAndModelCheckerBasedTestCaseGeneration}. 
$T$ satisfies Edge Coverage, when all edges (transitions) $e \in E \in \mathcal{G}$ are present in at least one $t \in T$. This criterion is commonly called \textit{0-Switch coverage} or \textit{All Transitions Coverage} \cite{CoverageCriteriaForStateTransitionTestingAndModelCheckerBasedTestCaseGeneration,Heimdahl2004TestSuiteReductionForModelBasedTests}. $T$ also satisfies \textit{Edge-Pair Coverage} when each path consisting of two adjacent edges $e \in E$ must occur at least once in at least one $t \in T$ \cite{ammann2016introduction,Li2009AnExperimentalComparisonOfFourUnitTestCriteria}. Edge-Pair Coverage is also mentioned in the literature as \textit{All Transition Pairs Coverage} and \textit{1-Switch Coverage} \cite{CoverageCriteriaForStateTransitionTestingAndModelCheckerBasedTestCaseGeneration}. Generalized \textit{$N$-Switch Coverage} is satisfied, when every combinations $N+1$ adjacent transitions (edges of $\mathcal{G}$) must occur at least once in a $t \in T$ \cite{ConcurrentNSwitchCoverageCriterionForGeneratingTestCasesFromPlaceTransitionNets}.

To generate $T$ from $\mathcal{G}$ or $\boldsymbol{\phi}$, a number of algorithms can be found in the literature. These algorithms differ by the test coverage criteria that are satisfied by the generated $T$. They also differ in the effectiveness of the generated $T$. Generally, two test sets that satisfy the same test coverage criteria may still differ in a number of test steps or test paths, which affects the overall effectiveness of the testing process. The algorithms were mostly implemented to generate test cases for specific classes of applications.  

Devroey~\textit{et~al.} proposed an algorithm to generate test suites for software product lines using Feature Diagram and Feature Transition System (FTS) as a SUT model \cite{AbstractTestCaseGenerationForBehaviouralTestingOfSoftwareProductLines}. Since FTS is a directed graph, this algorithm can also be applied to solve the problem discussed in this study. The algorithm uses a branch-and-bound approach and uses heuristics for efficient test path search. Instead of a breadth or depth-first search, the algorithm explores the graph using priorities, where the branch is prioritized when it has a higher score. In the algorithm, a score is used describing how this branch will visit many unvisited states, and how test sets generated by this branch will cover many new states. The score is evaluated using an accessibility matrix computed using a modified Warshall algorithm. Instead of distances, matrix cells contain feature expressions used to evaluate products capable of executing a transition, respectively, a set of transitions changing state of a system from one to another.

Another comparable algorithm was implemented by Alava \textit{et.~al.} \cite{AutomaticValidationOfJavaPageFlowsUsingModelBasedCoverageCriteria}. In their work, an approach is proposed to generate automated tests for \textit{Java Page Flow} web applications. The main input for this process is a directed graph called \textit{Design View (DView)}. DView is a directed graph, where nodes are pages or actions, and edges are links or forwards. The FSM of the page flows and the coverage criteria are obtained from DView. Comparable coverage criteria used in this approach are \textit{All pages} (equivalent to \textit{All Node Coverage}) and \textit{All actions} (equivalent to \textit{All Edge Coverage}). Test cases are generated to ensure the execution of these test paths.

Carvalho and Tsuchiya exploit model checking to generate test paths for SUT parts described as FSMs, as model checkers can generate counterexamples as proof when a model does not satisfy the specification \cite{CoverageCriteriaForStateTransitionTestingAndModelCheckerBasedTestCaseGeneration}. The tool uses the NuSMV modeling language to define an FSM. The method aims to support \textit{Node Coverage}, \textit{Edge Coverage}, and \textit{Edge-pair Coverage}. The coverage criteria were defined using the NuSMV language. When the SUT model and the test coverage criteria are prepared in NuSMV notation, the test paths are generated using a heuristic algorithm. Using their model, the authors identified this test set generation problem as an NP-hard covering problem. 

%%%%%%%%%%% Bestoun --- 28 November

Another comparable approach was proposed by Liu and Xu \cite{MTToolAToolForSoftwareModelingAndTestGeneration} to generate a test set for FSM \cite{MTToolAToolForSoftwareModelingAndTestGeneration}. RFSM is an extended FSM with a special label notation that gives the RFSM the ability to model more details, for example, a number of transitions or different types of node repetitions \cite{ANewApproachToGeneratingHighQualityTestCases}. This approach employs a Regular Finite State Machine (RFSM) to model the SUT using an MTTool tool graphical interface. An algorithm is used to transform the RFSM into ERE and then to generate the test paths. The used ERE is a classical regular expression for designing SUT behavior with extended grammar, giving the ability to model the nature of synchronous and concurrent task execution (transitions or sequences of transitions). ERE can be generated from RFSM. The SUT model in RFSM can be created in two ways: using an MTTool tool graphical editor or text input, using the commands of the author's proprietary R language. The ERE model is parsed into a set of submodels to avoid state space explosion problems during the generation of the test paths. The algorithm generating the test paths accepts this set of submodels and a set of test requirements, where these test requirements are parts of the SUT model or their combinations that the generated test paths must cover.

From the approaches using RE to model SUT behavior, Kilincceker~\textit{et al.} proposed a method that consists of the toolchain for the generation of test paths from SUT parts modeled as a regular expression or from an FSM, which is further conversed to RE \cite{Kilincceker2019}. In this approach, the context table is used during the generation of the test paths. Details can be derived from the toolchain source code available in a GitHub repository \cite{KilinccekerGithubMBIT4SW}. The tool is available freely for further analysis and comparison with newly developed alternatives.

Kilincceker~\textit{et al.} also presented an approach for generating test paths using a SUT specified in Hardware Description Language (HDL) language \cite{RegularExpressionBasedTestSequenceGenerationForHDLProgramValidation}. The work transfers HDL code into the FSM model that gave this approach a more expansive application field. During the generation of the test path, the FSM is further transformed to RE, in particular the extended RE model proposed by Liu et al. mentioned in \cite{AStudyForExtendedRegularExpression-basedTesting}. The approach also includes a model minimization process to speed up the generation of test paths. To obtain test paths, RE is parsed into a Syntax tree from which the test paths are finally generated using an algorithm specified in the study.

Fazli and Mohsen proposed the Strongly Connected Component (SCC) for the generation of prime paths and test sets based on them \cite{Fazli2019ATimeAndSpaceEfficientCompositionalMethodForPrimeAndTestPathsGeneration}. This method divides a problem of prime paths generation into smaller sub-problems that lead to better time and space efficiency. The input of this method is a Component Flow Graph, and its output is a set of test paths and a set of prime paths that have been covered. In their study, Fazli and Mohsen experimentally compare three approaches for the generation of prime paths for FSM. The results of the comparison showed that SCC performed well in terms of memory consumption and processing time.

Jia~\textit{et~al.} proposed a method for the generation of whole program paths to satisfy branch coverage \cite{Wei2018WholeProgramPathsGenerationMethod}. They employ a divide-and-conquer approach to achieve this goal, which makes this method similar to the SCC-based method discussed above. First, the authors generate a base path set (BPS) for each partial function of a Control Flow Graph, which serves as the first part of the SUT model. Then, in this graph, the algorithm identifies function call nodes using the second part of the SUT model, the Function Call Graph. These function call nodes are then joined with the test paths generated for a particular function call. This process is top-down to gradually join all function call nodes with function paths they call. Here, flags are used to mark functions that have actually been traversed. In this approach, recursive function calls are not supported, adding limitations to the method, since the recursive call is a common construct used routinely in programming.

From other alternatives, Klalil and Labiche presented an approach to generate FSM test paths, supported by a tool called STAGE-1 \cite{StateBasedTestsSuitesAutomaticGenerationTool}. Their test path generation method supports \textit{Round Trip Coverage}. Besides that, Random, Depth Traversal and Breadth Traversal criteria are discussed as alternatives. More test sets are produced for each given test coverage criterion (worst, best, and average cases) for further analysis.

Despite the fact that FSM testing is the well-established subarea of the system testing discipline, no work we have found so far is directly addressing: (1) the possibility to explicitly set a start and end of a test path in a SUT model and (2) to determine expected length range of the test paths. Regarding the modeling notation, no major rework or model redefinition is needed, and we easily build a SUT model for the proposed strategy by extension of $\mathcal{G}$ (see Section \ref{sec:proposed_sut_model} later). Regarding the test coverage criteria to address the goals of this paper, we need to define alternative criteria. The reason is to satisfy the first goal in which we need to neglect irrelevant or infeasible test paths (e.g., paths that are not starting and ending in explicitly given FSM states) to produce an effective set of test paths.

Considering the available algorithms, the majority of the approaches discussed in this Section, unfortunately, assume that a test path can start and end in any state of an FSM and does not provide a sufficient mechanism for expressing the required test path length. The available methods primarily focus on optimizing the test set. The goal is to minimize the number of test paths and steps while still satisfying the given test coverage criteria.

\section{The Proposed Approach}\label{sec:proposed_approach}

The \textbf{Flexible State Machine Test (FSMT)} strategy is an alternative approach to generate a more effective set of test paths to satisfy alternative test coverage criteria. 

FSMT is based on the following adjustments to the traditional \textit{N-switch Coverage} approach: (1) In addition to the start and end of the tested state machine, we also introduced the possible start and end of the test path, and (2) instead of sequences of uniform length implied by the N-switch Coverage criterion, we defined the length range of generated test paths. A consequence of these adjustments is that we need to define alternative test coverage criteria that must be satisfied by a set of test paths. We propose such criteria later in Section \ref{sec:test_coverage_criteria}.

\subsection{SUT model}
\label{sec:proposed_sut_model}

We model the SUT as a directed multigraph $\boldsymbol{G} = (V,E,L,\varepsilon,v_s,V_e,V_{ts},V_{te})$, where $V$ is a set of vertices representing FSM states, $E$ is a set of edges representing FSM transitions, and $L$ is a set of edge labels. Edge $e \in E$ defined by $\varepsilon:~E\rightarrow{}~\{(s, f, l)~|~s, f \in V \land l \in L\}$, where $s$ is the start vertex of edge $e$, $f$ is the end vertex of edge $e$, and $l$ is the label of edge $e$. Furthermore, $v_s \in V$ is the start vertex of the state machine, $V_e \subset V$ is a set of end vertices of the state machine, $V_{ts} \subset V$ is a set of possible start of test paths, $V_{te} \subset V$ is a set of possible end of test paths, $v_s \in V_{ts}$ and $V_{e} \subset V_{te}$. Moreover, $V_{ts}$ and $V_{te}$ can have nonempty intersect. During the creation of the SUT model, $V_{ts}$ and $V_{te}$ are defined by the test engineer using the design documentation of the SUT or his knowledge and experience with SUT testability.

The test path $p$ is a path in $\boldsymbol{G}$ that starts at $v_{ts} \in V_{ts}$ and ends at $v_{te} \in V_{te}$. A test path is a sequence of edges and $P$ is a set of all test paths.
The SUT model $\boldsymbol{G}$ is an input to the test path generation strategy defined later in Section \ref{sec:fsmt_strategy}, together with the test coverage criteria defined in Section \ref{sec:test_coverage_criteria}.

\subsection{Test coverage criteria}\label{sec:test_coverage_criteria}

Test coverage criteria serve to determine a level of guarantee, how many possible path combinations would be exercised by the paths present in a $P$. For this reason, the test coverage criterion is accepted as an input to an algorithm that generates $P$.

We use two test coverage criteria, \textit{FSMT-level-1 Coverage} and \textit{FSMT-level-2 Coverage}, which differ by the number of test path transitions. \textit{FSMT-level-1 Coverage} is designed for lower intensity tests and \textit{FSMT-level-2 Coverage} for higher intensity tests. This, in turn, added more flexibility to select what fits the testing goal in practice.

A set of all test paths $P$ satisfies \textit{FSMT-level-1 Coverage}, when all the following conditions are satisfied:

\begin{enumerate}
\item Each of the test paths $p \in P$ must start in a vertex from $V_{ts}$ and end in a vertex from $V_{te}$,

\item each vertex from $V_{ts}$ must be presented as the start vertex of a $p \in P$, and,

\item for each $p \in P$, $minLenght \leq length(p) \leq maxLength$, where $length(p)$ is the length of a test path $p$ in the number of its edges.   
\end{enumerate}

%\begin{enumerate}
%\item Each of test paths $p \in P$ must start in a vertex from $\{v_s\} \cup V_{ts}$ and end in a vertex from $V_e \cup V_{te}$,

%\item each vertex from $\{v_s\} \cup V_{ts}$ must be presented as a start vertex of a $p \in P$, and,

%\item for each $p \in P$, $minLenght \leq length(p) \leq maxLength$, where $length(p)$ is length of a test path $p$ in number of its edges.   
%\end{enumerate}

In addition to the rules given above, \textit{FSMT-level-1 Coverage} does not provide any additional requirement on how $V_{ts}$ and $V_{te}$ must be chained or combined in the test paths. Furthermore, it is not required that all vertices of $V_e \cup V_{te}$ be present as an end vertex of a $p \in P$. Also, \textit{FSMT-level-1 Coverage} in general does not require to visit either the entire $E \in \boldsymbol{G}$ or even $V \in \boldsymbol{G}$. The \textit{FSMT-level-1 Coverage} criterion is designed for lower intensity FSM tests when prioritization is needed for any reason, such as not having enough resources.

In the same way, a set of all test paths $P$ satisfies \textit{FSMT-level-2 Coverage}, when all the following conditions are satisfied:

\begin{enumerate}
\item $P$ satisfies \textit{FSMT-level-1 Coverage}, and,

\item each $e \in E \in \boldsymbol{G}$ that can be part of a $p \in P$ that starts at a vertex from $\{v_s\} \cup V_{ts}$, ends in a vertex from $V_e \cup V_{te}$ and $minLenght \leq length(p) \leq maxLength$, where $length(p)$ is the length of $p$ in the number of its edges, must be present in $p$.

\end{enumerate}

In this paper, we use a term \textit{subsume} to indicate that the meeting of a test coverage criterion $C_1$ subsumes $C_2$ if each test set that satisfies $C_1$ will also satisfy $C_2$. To this end, \textit{FSMT-level-2 Coverage} subsumes the \textit{FSMT-level-1 Coverage} criterion.

In contrast to \textit{FSMT-level-1 Coverage}, the \textit{FSMT-level-2 Coverage} is designed for more intensive tests when all FSM transitions must be executed during the tests. Still, \textit{FSMT-level-2 Coverage}, in general, does not require visiting neither $E \in \boldsymbol{G}$ nor $V \in \boldsymbol{G}$. Furthermore, a consequence of \textit{FSMT-level-1 Coverage} and \textit{FSMT-level-2 Coverage} that is defined in this way is that for certain $\boldsymbol{G}$ combined with certain ranges of $minLenght$ and $maxLenght$, $P$ it could not exist. In such a case, the problem can be solved by changing $minLenght$ and $maxLenght$, or adding more possible $V_{ts}$ and $V_{te}$ to $\boldsymbol{G}$.

\subsection{FSMT strategy}
\label{sec:fsmt_strategy}

The FSMT strategy comprises a few algorithms that aim to generate effective test sets. Here, the strategy is to generate a set of test paths for the SUT model $\boldsymbol{G}$, the expected length range of the test path, and the coverage criterion from the options given in \ref{sec:test_coverage_criteria}. To determine this test coverage criterion, we use a switch $testCoverage$ where its value 1 means \textit{FSMT-level-1 Coverage} and 2 means \textit{FSMT-level-2 Coverage}. For a certain test path length range $minLength - maxLength$, it is possible that $P$ would not meet the given test coverage criteria. In such a case, the test path length range $minLength - maxLength$ must be adjusted.

The main Algorithm \ref{alg:GenerateTestPaths} (\textit{GenerateTestPathsFSMT}) accepts the SUT model $\boldsymbol{G}$ (defined in Section \ref{sec:proposed_sut_model}), minimal length of test paths (denoted as $minLength$), maximum length of test paths (denoted as $maxLength$), and a switch for the test coverage criterion ($testCoverage$, defined in Section \ref{sec:test_coverage_criteria}). The algorithm returns a set of test paths $P$ and a set of uncovered edges $E_{uncovered}$. First, the algorithm iterates at all vertices of $\boldsymbol{G}$ in which a test path can start (denoted as $V_{ts}$) and tries to find the shortest path in the range to a vertex in which a test path can end (denoted as $V_{te}$). $V_{ts}$ and $V_{te}$ are part of the SUT model $\boldsymbol{G}$, which is given to the Algorithm \ref{alg:GenerateTestPaths} as input. In this iterating, the Algorithm \ref{alg:FindShortestPathInRange} (\textit{FindShortestPathInRange}) is used. Edges used in the paths found in this process are considered covered. The uncovered edges, denoted as $E_{uncovered}$, are those that are not part of any of these identified paths. After that, if $testCoverage = 2$, the algorithm tries to satisfy the \textit{FSMT-level-2} criterion. In this case, it is taking random edges from the set of uncovered edges and tries to find the shortest path that is (1) longer than $minLength$ (inclusive), and (2) shorter than $maxLength$ (inclusive), and (3) composing of the maximum number of uncovered edges, using Algorithm \ref{alg:FindShortestPathInRangeForEdge} (\textit{FindShortestPathInRangeForEdge}).

The Algorithm \ref{alg:FindShortestPathInRange} (\textit{FindShortestPathInRange}) accepts the SUT model $\boldsymbol{G}$, minimal length of test paths ($minLength$), maximal length of test paths ($maxLength$), $testCoverage$ switch, set of uncovered edges ($E_{uncovered}$) and a vertex in which the algorithm starts construction of a test path (denoted as $v_{ts}$). The Output of the Algorithm \ref{alg:FindShortestPathInRange} is a set of test paths $P$. At the beginning, the next path to proceed (denoted $p_{next}$) is set to an empty path. The start vertex of the constructed path (denoted as $v_{last}$) is set to $v_{ts}$ and the queue of paths to process $\boldsymbol{Q}$ is initiated empty. After that, there is a cycle repeated while $p_{next}$ is not empty. At the beginning of this cycle, the algorithm tests if the size of $p_{next}$ is the same as or lower than the maximal length of the test path ($maxLen$). If so, the algorithm checks if the constructed path ends at one of the nodes of $V_{ts}$. If this condition is met, the algorithm considers this path a test path to be returned. Otherwise, it will check if the length of the constructed path is less than $maxLen$. If this is true, the algorithm creates a new path $p_{new}$ for each outgoing edge from $v_{next}$ by concatenating this edge with $p_{next}$. This part is done using the Algorithm \ref{alg:RemoveParallelEdges} (\textit{RemoveParallelEdges}). Then $p_{new}$ is pushed to the queue $\boldsymbol{Q}$ which contains paths to process. At the end of the cycle, if $\boldsymbol{Q}$ is not empty, the algorithm pulls another path to be processed from $\boldsymbol{Q}$, assigns it to $p_{next}$ and sets $v_{next}$ to the last vertex of $p_{next}$, otherwise $p_{next}$ is set to be an empty path. At the end of the cycle, there is no test path composed, so an empty path is returned.

%%%%%%%%%%% Bestoun --- 30 November

The algorithm \ref{alg:RemoveParallelEdges} (\textit{RemoveParallelEdges}) accepts a set of edges from which parallel edges must be removed (denoted as $E_{toFilter}$), a set of uncovered edges (denoted as $E_{uncovered}$) and  $testCoverage$ switch. The output of this algorithm is a set of edges in which no parallel edges occur, denoted as $E_{filtered}$. The algorithm iterates over $E_{toFilter}$, and the actual iterated edge is denoted $e_{unfiltered}$. The algorithm tries to find a parallel edge $e_{parallel}$ to the actual edge $e_{unfiltered}$ in the set $E_{filtered}$. If $e_{parallel}$ does not exist, it will add $e_{unfiltered}$ to the set $E_{filtered}$. Otherwise, if all edges have to be covered (as indicated by $testCoverage$ switch) and if $e_{parallel} \notin E_{uncovered}$, the algorithm swaps $e_{parallel} \in E_{filtered}$ with the actual edge $e_{unfiltered}$. Finally, the algorithm returns set $E_{filtered}$.

The algorithm \ref{alg:FindShortestPathInRangeForEdge} (\textit{FindShortestPathInRangeForEdge}) accepts an edge $e_{uncovered}$ that must be present on a built test path, the SUT model $\boldsymbol{G}$, $minLen$, $maxLen$, $testCoverage$ and a set of uncovered edges $E_{uncovered}$. The algorithm starts its exploration in $e_{uncovered}$ and from this edge, it traverses the graph $\boldsymbol{G}$ forwards (following the edges directions) to an end vertex from $V_{te}$ of a possible test path. The algorithm then traverses the $\boldsymbol{G}$ backwards (in reverse direction than the directions of the edges) to a start vertex from $V_{ts}$ of a possible test path. The goal is to find a test path that starts at a vertex of $V_{te}$, ends at a vertex of $V_{ts}$, contains $e_{uncovered}$, is longer than $minLen$ inclusive and is shorter than $maxLen$ inclusive. If no such path exists, an empty path is returned. This exploration is done by the Breadth First Search (BFS) principle simultaneously in both discussed directions and is described in a technical subroutine specified in Algorithm \ref{alg:FindPathInRangeForEdgeDirected} (\textit{FindPathInRangeForEdgeDirected}).

%Algorithm \ref{alg:FindPathInRangeForEdgeDirected} analyzes if a path given in a queue $Q_{processed}$ is a suitable candidate for a test path in the first exploration direction. If so, it tries to construct a full test path using a path given in $Q_{processed}$ in the opposite direction. When a full test path is constructed, it is returned, and the algorithm ends. Otherwise, if there are no suitable candidates in $Q_{processed}$, the actual candidate from the first exploration direction is stored in a cache $M_{other}$ to be analyzed later.%

Algorithm \ref{alg:FindPathInRangeForEdgeDirected} takes the next path to be processed, checks whether it is possible to use this path to create the full test path, (in the current or later iteration) and initiates preparation of the next moves that will be processed by this algorithm in the next iterations. The algorithm \ref{alg:FindPathInRangeForEdgeDirected} uses two sub-routines, \textit{EvaluateCandidate}, specified in Algorithm \ref{alg:EvaluateCandidate} and \textit{PrepareNextMoves}, specified in Algorithm \ref{alg:PrepareNextMoves}.

The Sub-routine \textit{EvaluateCandidate} (Algorithm \ref{alg:EvaluateCandidate}) accepts a semi-test path and evaluates whether this path can be used altogether with an actual found semi-path for construction of the full test path. If it does so, it returns this full test path. Otherwise, it stores the evaluated semi-test path for possible later usage. The Sub-routine \textit{PrepareNextMoves} (Algorithm \ref{alg:PrepareNextMoves}) accepts a path, extends this path with the next step in an appropriate direction, and puts this path in a queue of paths that are stored for further processing.

%%%%%%%%%%% Bestoun --- 30 November

\begin{algorithm*}
\caption{Generate test paths for SUT model by FSMT strategy}\label{alg:GenerateTestPaths}
%\hspace*{\algorithmicindent} \textbf{\bf Function: GenerateTestPathsFSMT}\\
\begin{flushleft}
\textbf{\bf Function: GenerateTestPathsFSMT}\\
%\hspace*{\algorithmicindent} \textbf{Input:} SUT model $\boldsymbol{G}$, $minLength, maxLength, testCoverage$\\
\textbf{Input:} $\boldsymbol{G}$, $minLength, maxLength, testCoverage$\\
%\hspace*{\algorithmicindent} \textbf{Output:} Set of test paths $P$ and set of uncovered edges $E_{uncovered}$
\textbf{Output:} Set of test paths $P$ and set of uncovered edges $E_{uncovered}$
\end{flushleft}
\begin{algorithmic}[1]
\State $P \gets \emptyset$  \Comment{an empty set of test paths}
\State $E_{uncovered} \gets E$ \Comment{a set of edges uncovered by test paths}

\ForEach {$v_{ts} \in V_{ts}$}

    \State \raggedright $p_{new} \gets $
    \textbf{FindShortestPathInRange(}$\boldsymbol{G},$ $minLength,$ $maxLength,$
    $testCoverage,$ $E_{uncovered},$ $v_{ts}$\textbf{)}
    
    \If{$p_{new}$ is not empty}
        \State $E_{uncovered} \gets E_{uncovered} \setminus \{ e~|~e$ is present in $p_{new}\}$
        \State $P \gets P \cup \{p_{new}\}$
    \EndIf
\EndFor
\If{ $testCoverage = 2$ }
    \While {$E_{uncovered}$ is not empty}

        \State $e_{uncovered} \gets$ any $e \in E_{uncovered}$
        \State $E_{uncovered} \gets E_{uncovered} \setminus \{e_{uncovered}\}$
        \State $p_{new} \gets$ \textbf{FindShortestPathInRangeForEdge(}$e_{uncovered},$ $\boldsymbol{G},$ $minLength$,
        %\par\hskip\algorithmicindent 
        $maxLength$, $testCoverage,$ $E_{uncovered}$\textbf{)}
        \If{$p_{new}$ is not empty}
            \State $E_{uncovered} \gets E_{uncovered} \setminus \{ e~|~e$ is present in $p_{new}\}$
            \State $P \gets P \cup \{p_{new}\}$
        \EndIf
    \EndWhile
\EndIf
\State \textbf{return} ($P$,$E_{uncovered}$) \Comment{return all found paths and a set of uncovered edges}
%\Statex{\bf end function}
\end{algorithmic}
\end{algorithm*}

\begin{algorithm*}
\caption{Find the shortest path in range}\label{alg:FindShortestPathInRange}
\begin{flushleft}
\textbf{Function: FindShortestPathInRange}\\
\textbf{Input:} SUT model $\boldsymbol{G}, minLength, maxLength, testCoverage, E_{uncovered}, v_{ts}$\\
\textbf{Output:} Path $p$
\Comment{if no path is found then an empty path is returned}
\end{flushleft}
\begin{algorithmic}[1]

\State $p_{next} \gets$ empty path
\State $v_{last} \gets v_{ts}$
\State $\boldsymbol{Q}$ is an empty queue of paths
\Do
    \If{$|p_{next}| \leq maxLength$}
        \If{($|p_{next}| \geq minLength$) $\land$ ($v_{last} \in V_{te}$)}
            \State \textbf{return} $p_{next}$
        \EndIf
        \If{$p_{next} < maxLength$}
            \State $E_{outgoing} \gets$ edges outgoing from $v_{last}$
            \State \raggedright $E_{outgoing} \gets $ \textbf{RemoveParallelEdges(} $E_{outgoing}, E_{uncovered},$ $testCoverage$ \textbf{)} 
            
            \ForEach{$e_{outgoing} \in E_{outgoing}$}
                \State $p_{new} \gets p_{next}$ appended with $e_{incoming}$ at its end
            
                \State $\textbf{push}$ $p_{new}$ to $\boldsymbol{Q}$
            \EndFor
        \EndIf
    \EndIf
    \If{$\boldsymbol{Q}$ is not empty}
        \State $p_{next} \gets$ \textbf{pull} from $\boldsymbol{Q}$
        \State $v_{next} \gets$ the last vertex of $p_{next}$
    \Else
        \State $p_{next} \gets$ empty path
    \EndIf
\doWhile{$p_{next}$ is not empty}
\State \textbf{return} empty path\Comment{no path found}
\end{algorithmic}
\end{algorithm*}

\begin{algorithm}
\caption{Remove parallel edges}\label{alg:RemoveParallelEdges}
\begin{flushleft}
\textbf{Function: RemoveParallelEdges}\\
\textbf{Input:} %$E_{to\char`_filter}, E_{uncovered}, testCoverage$\\
$E_{toFilter}, E_{uncovered}, testCoverage$\\
\textbf{Output:} Set of edges $E_{filtered}$
\end{flushleft}
\begin{algorithmic}[1]
%\Statex{\bf function RemoveParallelEdges}
\State $E_{filtered} \gets \emptyset$  \Comment{an empty set of edges}
\ForEach{$e_{unfiltered} \in E_{toFilter} $}
    \State $e_{parallel} \gets$ an edge from  $E_{filtered}$ parallel to $e_{unfiltered}$, if such an edge does not exist, $e_{parallel} \gets nil$
    \If{$e_{parallel}$ is $nil$}
        \State $E_{filtered} \gets E_{filtered} \cup \{ e_{unfiltered} \}$
    \ElsIf{$testCoverage = 2$}
        \If{$e_{parallel} \notin E_{uncovered}$}
            \State $E_{filtered} \gets ( \, E_{filtered} \setminus \{ e_{parallel} \}  ) \, \cup \{ e_{unfiltered} \}$
        \EndIf
    \EndIf
\EndFor
\end{algorithmic}
\end{algorithm}

\begin{algorithm*}
\caption{Find the shortest path in the range of the edge}\label{alg:FindShortestPathInRangeForEdge}
\begin{flushleft}
\textbf{Function: FindShortestPathInRangeForEdge}\\
\textbf{Input:} $e_{uncovered}, \boldsymbol{G}, minLength, maxLength, testCoverage, E_{uncovered}$\\
\textbf{Output:} Path $p$
\Comment{if no path is found then an empty path is returned}
\end{flushleft}
\begin{algorithmic}[1]
%\Statex{\bf function FindShortestPathInRangeForEdge}
%    \Comment{all followed maps are mapped by path length}
%    \State $E_{map} \gets$ an empty map of paths to some end node
%    \State $S_{map} \gets$ an empty map of paths to some start node
%    \State $E_{queue} \gets$ an empty queue of paths to some end node
%    \State $S_{queue} \gets$ an empty queue of paths to some start node
    %\State  are indexed by path length

    \State $E_{map}$ and $S_{map}$ are empty maps of paths. The key in the map is the length of the path. 
    \State $E_{queue}$ and $S_{queue}$ are empty queues of paths
    \State $p_{end}$ and $p_{start}$ are paths one edge long created from ${e_{uncovered}}$
    \State \textbf{push } $p_{end}$ to $E_{queue}$
    \State \textbf{push } $p_{start}$ to $S_{queue}$

    \State $startMin \gets 1$, 
    $endMin \gets 1$,
    $startMinCount \gets 1$,
    $endMinCount \gets 1$
    \State $startMaxCount \gets 0$, 
    $endMaxCount \gets 0$
    
    \While{($E_{queue}$ is not empty) $\lor$ ($S_{queue}$ is not empty)}

        \If{$S_{queue}$ is not empty}

        \State \raggedright ($p, startMin, startMinCount, startMaxCount, S_{queue}, S_{map}) \gets$ \textbf{FindPathInRangeForEdgeDirected(} $\textit{minLength, maxLength, testCoverage, startMin, endMin, startMinCount, startMaxCount,} S_{map}, E_{map}, S_{queue}, V_{ts}, TRUE$ \textbf{)}
            \If{$p$ is not empty} \textbf{return} $p$
            \EndIf
        \EndIf
        \If{$E_{queue}$ is not empty}
            \State ($p, endMin, endMinCount, endMaxCount, E_{queue}, E_{map}) 
            \gets$ \textbf{FindPathInRangeForEdgeDirected(} $\textit{minLength, maxLength, testCoverage, endMin, startMin, endMinCount, endMaxCount,} E_{map}, S_{map}, E_{queue}, V_{te}, FALSE$ \textbf{)}
            \If{$p$ is not empty} \textbf{return} $p$
            \EndIf
        \EndIf
    \EndWhile
\State \textbf{return} empty path
\end{algorithmic}
\end{algorithm*}

\begin{algorithm*}
\caption{Find the path in the range for the edge}\label{alg:FindPathInRangeForEdgeDirected}
\begin{flushleft}
\textbf{Function: FindPathInRangeForEdgeDirected}\\
\textbf{Input:} $minLength, maxLength, testCoverage, processedMin, otherMin,$
$processedMinCount, processedMaxCount, M_{processed}, M_{other},Q_{processed},$ $V_{destination},$ $backward$\\
\textbf{Output:} path $p,$ $processedMin,$ $processedMinCount,$ $processedMaxCount,$ $Q_{processed},$ $M_{processed}$
\Comment{if no path is found then an empty path is returned}
\end{flushleft}
\begin{algorithmic}[1]
\State $p_{processed} \gets$ \textbf{pull} from $Q_{processed}$, decrease $processedMinCount$ by 1
%\State $processedMinCount \gets processedMinCount - 1$
\If{$|p_{processed}| \leq  maxLength$}
    \If{$backward$}
        \State $v_{processed} \gets$ start vertex of path $p_{processed}$
    \Else
        \State $v_{processed} \gets$ end vertex of path $p_{processed}$
    \EndIf
    \If{$v_{processed} \in V_{destination}$}
        \Comment{Destination vertex reached, actual path will be evaluated as a candidate to create full test path}

        \State \raggedright ($p_{full}, M_{processed}) \gets$ \textbf{EvaluateCandidate(}~$p_{processed},$ $M_{processed},$ $M_{other},$ $backward,$ $otherMin,$ $maxLength$ \textbf{)}
        \If{$p_{full}$ is not an empty path}
            \State \raggedright \textbf{return} ($p_{full},$ $processedMin,$ $processedMinCount,$ $processedMaxCount,$ $Q_{processed},$ $M_{processed}$)
        \EndIf
    \EndIf
    \If{$(|p_{processed}| + 1 + otherMin) < maxLength$}
        \State $(Q_{processed},~processedMaxCount) \gets$ \raggedright \textbf{PrepareNextMoves(} $Q_{processed}$, $p_{processed}$, $v_{processed}$, $processedMaxCount$ \textbf{)}
    \EndIf
\EndIf

\If{$proccesedMinCount = 0$}
    \State $startMinCount \gets startMaxCount$, 
    $startMaxCount \gets 0$
    \State $processedMin \gets processedMin + 1$
\EndIf
\State \textbf{return} 
(empty path, $processedMin,$ $processedMinCount,$ $processedMaxCount,$ $Q_{processed},$ $M_{processed}$)

\end{algorithmic}
\end{algorithm*}

\begin{algorithm*}

\caption{Evaluate candidate}\label{alg:EvaluateCandidate}
\begin{flushleft}
\textbf{Function: EvaluateCandidate}\\
\textbf{Input:} $p_{processed}, M_{processed}, M_{other}, backward, otherMin, maxLength$ \\
\textbf{Output:} path $p,$ $M_{processed}$
\Comment{if no path is found then an empty path is returned}
\end{flushleft}
\begin{algorithmic}[1]

\State $lowerBound \gets$ max(~$0,~minLength - |p_{processed}|$ )$ + 1$
\State $upperBound \gets maxLength - |p_{processed}| + 1$
\ForEach{$i \in \{ lowerBound,~...~, upperBound \} $} %\Comment{iterate by 1 from $lowerBound$ to $upperBound$}
    \If{$M_{other}$~contains key $i$}
        \State $p_{other} \gets M_{other}[ \, i ] \,$  \Comment{value for key $i$}
        \If{$backward$}
            \State $p \gets
            %(p_{processed}$ without its last edge) with $p_{other}$ added at its end
            (p_{processed}$ without its last edge) appended with $p_{other}$
        \Else
            \State $p \gets
            %( p_{other}$ without its last edge) with $p_{processed}$ added at its end
            ( p_{other}$ without its last edge) appended with $p_{processed}$
        \EndIf
        \State \raggedright \textbf{return} ($p,$ $M_{processed}$)
    \EndIf
\EndFor
\If{$(|p_{processed}| + otherMin) \leq maxLength$}
    \If{$M_{processed}$ does not contain key $|p_{processed}|$}
        \State $M_{processed}[ \,|p_{processed}|~] \, \gets p_{processed}$ 
%                \Comment{set a value for key |p_{processed}|}
    \ElsIf{($testCoverage=2)~\land~(M_{processed}[ \,|p_{processed}|~]$ contains less edges from $E_{uncovered}$ than $p_{processed}$) \,}          
        \State $M_{processed}[ \,|p_{processed}|~] \, \gets p_{processed}$
    \EndIf
\EndIf

\State \textbf{return} (empty path, $M_{processed}$)

\end{algorithmic}

\end{algorithm*}

\begin{algorithm*}

\caption{Prepare next moves}\label{alg:PrepareNextMoves}
\begin{flushleft}
\textbf{Function: PrepareNextMoves}\\
\textbf{Input:} $Q_{processed}$, $p_{processed}$, $v_{processed}$, $processedMaxCount$\\
\textbf{Output:} $Q_{processed}$, $processedMaxCount$
\Comment{if no path is found then an empty path is returned}
\end{flushleft}
\begin{algorithmic}[1]

%\State $E_{incoming} \gets  \{ e | e \in \boldsymbol{E}: end~vertex~of~e~is~the~same~as~v_{first}\}$
\If{$backward$}
    \State $E_{next} \gets $ all edges incoming to $v_{processed}$
\Else
    \State $E_{next} \gets $ all edges outgoing from $v_{processed}$
\EndIf

\State $E_{next} \gets $ \textbf{RemoveParallelEdges(}~$E_{next}, E_{uncovered}, testCoverage$ \textbf{)} 

\ForEach{$e_{next} \in E_{next} $}
    % CONTINUE FROM HERE VASEK
    \If{$backward$}
        \State $p_{new} \gets e_{next}$ added at the start of $p_{processed}$ 
    \Else
        \State $p_{new} \gets           p_{processed}$ with $e_{next}$ added at its end
    \EndIf
    
    \State \textbf{push}~$p_{new}$~to~$Q_{processed}$
    \State $processedMaxCount \gets processedMaxCount + 1$
\EndFor

\State \textbf{return} ($Q_{processed}$, $processedMaxCount$)

\end{algorithmic}

\end{algorithm*}

\section{N-switch set reduction strategy} \label{sec:brute_force_approach}

To have a comparable alternative to the proposed FSMT, in the initial experiments, we use \textbf{N-switch Set Reduction (NSR) strategy}. It is based on the generation of all \textit{N-Switch Coverage} test paths and subsequent filtering of these paths. There are two levels of filtering done:

\begin{enumerate}
\item Remove the paths that do not start at a vertex from $\{v_s\} \cup V_{ts}$ and end at a vertex from $V_e \cup V_{te}$, and,
\item remove further duplication in the test paths that can be removed from $P$, so that $P$ still satisfies the test coverage criteria defined in Section \ref{sec:test_coverage_criteria}. 
\end{enumerate}

The strategy for the second filtering level differs depending on the test coverage criteria.

The main algorithm \ref{alg:BruteForceApproachTestPathsGenerator} (\textbf{GenerateTestPathsNSR}) accepts the SUT model $\boldsymbol{G}$, minimal length of test paths ($minLength$), maximal length of the test paths ($maxLength$) and a switch for the test coverage criterion ($testCoverage$). The algorithm first generates all paths in $\boldsymbol{G}$ of length $N$, where $minLength \leq N \leq maxLength$. This is done by a subroutine described in algorithm \ref{alg:FindPathsInRangeForEdgeRecursive} (\textbf{FindPathsInRange- ForEdgeRecursive}) that iterates over all $\boldsymbol{G}$ edges. In each iteration, the algorithm generates all possible paths of the required length beginning at the start vertex of the iterated edge. This job is done recursively. In one iteration, Algorithm \ref{alg:FindPathsInRangeForEdgeRecursive} checks if the path has the required length. If so, it puts it in the set of results $P$. After that, the algorithm checks if this path can be extended. If this condition is met, the algorithm calls itself recursively for each outgoing edge of the path the last vertex until this exploration is within the given test path limit ($minLength$ to $maxLength$).

After all possible paths of length $N$ are generated, algorithm \ref{alg:FilterTestPaths} (\textbf{FilterTestPaths}) reduces $P$ to keep only paths that are valid test paths from the \textit{FSMT-level-1} and \textit{FSMT-level-2} viewpoint - the test path starts in $\{v_s\} \cup V_{ts}$ and ends at a vertex from $V_e \cup V_{te}$. In this phase, $P$ still contains a lot of duplication in the test path. Therefore, another reduction $P$ is performed using Algorithm \ref{alg:ReduceTestPathsSet} (\textbf{ReduceTestPathsSet}). Here,the  paths of $p$ are analyzed if more paths start in a particular vertex from $\{v_s\} \cup V_{ts}$ and if so, only one of these paths is kept in $P$.

%%%%%%%%%%% Bestoun --- 30 November

\begin{algorithm}
\caption{Generate test paths for the SUT model by NSR strategy}
\label{alg:BruteForceApproachTestPathsGenerator}
\begin{flushleft}
\textbf{Function: GenerateTestPathsNSR}\\
\textbf{Input:} SUT model $\boldsymbol{G},minLen,maxLen,testCoverage$\\
\textbf{Output:} Set of test paths $P$
\end{flushleft}
\begin{algorithmic}[1]
\State $P \gets \emptyset$ \Comment{empty set of paths}
\ForEach{$e \in E$}
%    \State $P \gets \emptyset$ \Comment{empty set of paths}
    \State $p \gets$ empty path
    \State \raggedright $P_{new} \gets$ \textbf{FindPathsInRangeForEdgeRecursive(}~$p,P,e,\boldsymbol{G},minLen,maxLen$ \textbf{)}
    \State $P \gets$ $P \cup P_{new}$
\EndFor
\State $P \gets$ \textbf{FilterTestPaths(}$P,\boldsymbol{G}$ \textbf{)}
\State $P \gets$ \textbf{ReduceTestPaths(}$P,\boldsymbol{G},testCoverage$ \textbf{)}

\State \textbf{return} $P$
\end{algorithmic}
\end{algorithm}

\begin{algorithm}
\caption{Find paths in range for edge}
\begin{flushleft}
\label{alg:FindPathsInRangeForEdgeRecursive}
\textbf{Function: FindPathsInRangeForEdgeRecursive}\\
\textbf{Input:} $p,P,e,\boldsymbol{G},minLen,maxLen$\\
\textbf{Output:} Set of test paths $P$
\end{flushleft}
\begin{algorithmic}[1]
    \If{($|p| \geq minLen \land |p| \leq maxLen$)}
        \State $P \gets$ $P \cup \{p\}$
    \EndIf
    \If{$|p| < maxLen$}
        \State $p \gets$ $p$ with $e$ added at its end
        \State $E_{outgoing} \gets$ edges outgoing of $v$, $v$ is a vertex to which $e$ is incoming
        \ForEach{$e_{outgoing} \in E_{outgoing}$}
            \State $P_{new} \gets$ \textbf{FindPathsInRangeForEdgeRecursive(}$p,$ $P,$ $e,$ $\boldsymbol{G},$ $minLen,$ $maxLen$ \textbf{)}
            \State $P \gets$ $P \cup P_{new}$
        \EndFor
    \EndIf
\State \textbf{return} $P$
\end{algorithmic}
\end{algorithm}

\begin{algorithm}
\caption{Filter test paths}
\label{alg:FilterTestPaths}
\begin{flushleft}
\textbf{Function: FilterTestPaths}\\
\textbf{Input:} $P,\boldsymbol{G}$\\
\textbf{Output:} Set of test paths $P_{filtered}$ 
\end{flushleft}
\begin{algorithmic}[1]
    \State $P_{filtered} \gets \emptyset$ \Comment{empty set of paths}
    \ForEach{$p \in P$}
        \State $v_s \gets$ the first vertex of $p$, $v_e \gets$ the last vertex of $p$
        \If{($v_s \in V_{ts} \land v_e \in V_{te}$)}
            \State $P_{filtered} \gets P_{filtered} \cup \{p\}$
        \EndIf
    \EndFor
    \State \textbf{return} $P_{filtered}$
\end{algorithmic}
\end{algorithm}

\begin{algorithm}
\caption{Reduce test paths set}
\label{alg:ReduceTestPathsSet}
\begin{flushleft}
\textbf{Function: ReduceTestPathsSet}
\textbf{Input:} $P,\boldsymbol{G},testCoverage$\\
\textbf{Output:} Set of test paths $P_{filtered}$ 
\end{flushleft}
\begin{algorithmic}[1]
    \State $P_{reduced} \gets \emptyset$ \Comment{empty set of paths}
%    \State $S_{covered} \gets \emptyset$, $E_{covered} \gets \emptyset$  \Comment{empty sets of vertices}
    \State $S_{covered} \gets \emptyset$, \Comment{empty set of vertices}
    \State $E_{coveredEdges} \gets \emptyset$ \Comment{empty set of edges}
    \ForEach{$p \in P$}
%        \State $v_s \gets$ the first vertex of $p$, $v_e \gets$ the last vertex of $p$
        \State $v_s \gets$ the first vertex of $p$
        \If{($v_s \notin V_{ts} \lor ((testCoverage=2)~\land $ an edge of $p$ is not in $E_{coveredEdges})$)} 
%        \If{($v_s \notin V_{ts} \lor v_e \notin V_{te} \lor ((testCoverage=2)~\land $ some edge of $p$ is not in $E_{coveredEdges})$)}
            \State $P_{reduced} \gets P_{reduced} \cup \{p\}$
            \State $S_{covered} \gets S_{covered} \cup \{v_s\}$
%            \State $S_{covered} \gets S_{covered} \cup \{v_s\}$, $E_{covered} \gets E_{covered} \cup \{v_e\}$
            
            \State $E_{coveredEdges} \gets E_{coveredEdges} \cup \{$edges of $p\}$
        \EndIf
    \EndFor
    \State \textbf{return} $P_{reduced}$
\end{algorithmic}
\end{algorithm}

\section{Initial implementation of the proposed strategy}
\label{ref:initial_implementation_of_the_proposed_strategy}

We have implemented the FSMT strategy on the Oxygen experimental MBT platform, developed by our research group \cite{bures2015pctgen,bures2017prioritized}. The Oxygen platform is implemented in Java and can be downloaded and run as an executable JAR file. Java 1.8 Standard Development Kit or Java Runtime 1.8 environment is required to be installed on a local machine. The Oxygen platform with FSTM\footnote{http://still.felk.cvut.cz/download/oxygen3.zip} has been released for free public use. Oxygen provides a visual editor to create an SUT model $\boldsymbol{G}$. The schema is based on a simplified UML notation for state machines. Since the possible test paths start and end in UML are not available, they are marked by the color filling of a particular state symbol. The FSM states that are in $V_{ts}$ are marked by a green background, the states in $V_{te}$ by a red background, and if a state belongs to both $V_{ts}$ and $V_{te}$, yellow coloring is used.

%\begin{figure}[hbt!]
%    \centering
%    \includegraphics[width=8cm]{figures/oxygen_screenshot_model.png}
%    \caption{SUT model creation in the Oxygen platform.}
%    \label{fig:oxygen_screenshot_model}
%\end{figure}

The start and end states of the FSM and its states are dragged to a canvas from the upper panel. If a state belongs to $V_{ts}$ or $V_{te}$, it can be selected by a checkbox on the right panel when a particular state is selected in a diagram. FSM transitions are created by dragging a mouse from one state to another. By default, nodes are marked by letters, and transitions are marked by numbers when first placed on a canvas. These names can be changed in the right panel when a particular object is selected to edit. Other metadata such as state or transition description or test step expected result could be added there as well. 

\begin{figure*}[hbt!]
    \centering
    \includegraphics[width=13cm]{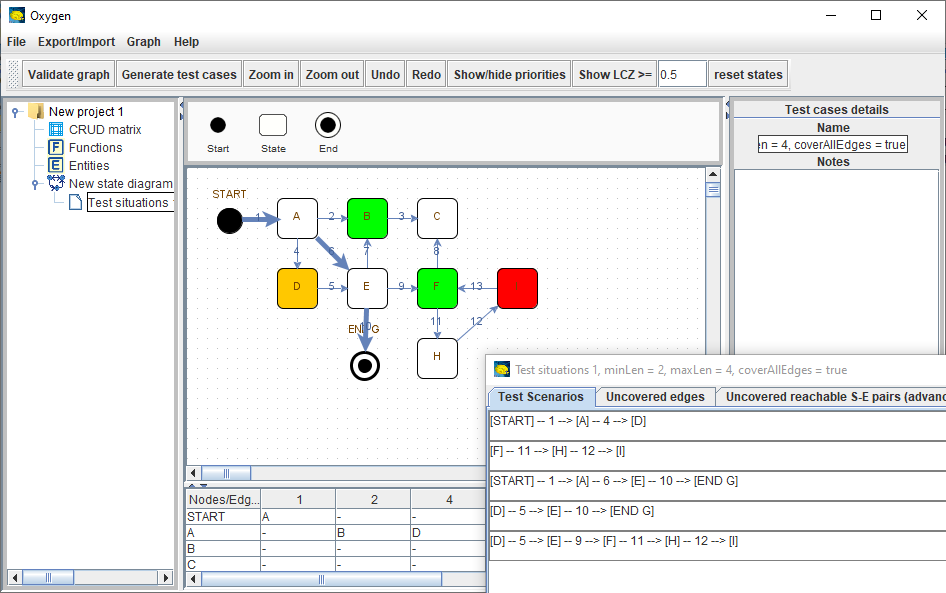}
    \caption{Visualization of generated test paths in the Oxygen platform.}
    \label{fig:oxygen_screenshot_test_paths}
\end{figure*}

The created FSM can be validated for basic modeling errors such as inaccessible states, missing start, and others. When a schema is valid for FSM, the implemented FSMT strategy can generate the test paths. At this stage, the parameters $minLength$, $maxLength$, and $testCoverage$ are entered into a dialog box. More test sets with different parameters can be generated and are stored in the project tree in the left application panel. From this project tree, the test set can be opened in a separate window and selected test paths can be visualized in the SUT model by a bold line (see Figure \ref{fig:oxygen_screenshot_test_paths}).

The generated test paths can then be exported in open formats based on XML, CSV, and JSON. The exported files can be easily used by a test management tool that supports a manual testing process or a test automation tool. In the same way, we implemented the NSR strategy, which works as a baseline to compare FSMT within the initial experiments.

%%%%%%%%%%% Bestoun --- 30 November

\section{Evaluation Experiments}
\label{sec:initial_experiments}

Our FSMT strategy is unique in concurrently addressing the need for explicitly defined test paths that start and end and address the possibility of specifying the expected length of the test paths. To this end, it is challenging to identify a state-of-the-art strategy that would be completely comparable. So far, the NSR approach presented in Section \ref{sec:brute_force_approach} is the best comparable option for the proposed FSMT strategy. The FSMT has been successfully applied in Skoda Auto car manufacturer to integration and acceptance tests of the produced automobiles. Due to the non-disclosure agreement, we are not allowed to give extensive details; however, industrial FSMs from this project were used as problem instances in the following experiments.

\subsection{Experiment method and set up}\label{sec:experiment_set_up}

In the following experiments, we used the FSMT and NSR implemented on the Oxygen platform to generate $P$ for the $\boldsymbol{G}$ problem instances described in Section \ref{sec:problem_instances}. We run the FSMT and NSR for four sets of length ranges (determined by the intervals $minLength$ to $maxLength$), as specified in Table \ref{tab:experiment_length_ranges}.

\begin{table*}%[h!]
\centering
\caption{Allowed test path length ranges for the experiments. LR stands for Length Range.}
%\begin{adjustbox}{max width=\textwidth}
\begin{tabular}{|c|c|c|c|c|}

\hline
\rowcolor{gray!25} \textbf{LR set ID} & $minLength$ & $maxLength$ & \textbf{variability in test path length}\\
\hline
\cellcolor{gray!25} \textbf{1} & 2 & 4 & 2\\
\hline
\cellcolor{gray!25} \textbf{2} & 2 & 6 & 4\\
\hline
\cellcolor{gray!25} \textbf{3} & 2 & 8 & 6\\
\hline
\cellcolor{gray!25} \textbf{4} & 4 & 8 & 4\\
\hline
\end{tabular}
%\end{adjustbox}

\label{tab:experiment_length_ranges}
\end{table*}

We analyze the following properties of the generated test paths:
\begin{itemize}
\item $|P|$
\item $len$ = total length of all $p \in P$, measured in the number of edges
\item $avlen$ = average length of all $p \in P$, measured in number of edges
\item $unique$ = number of unique edges in all $p \in P$
\item $ut = \frac{len}{unique}$  
\end{itemize}

The $ut$ defined above expresses how many non-unique FSM transitions ($\boldsymbol{G}$ edges) need to repeat in a test path to test all unique transitions. The higher $ut$ is, the higher this "edge duplication" is in a test set $P$.

\begin{figure}%[hbt!]
    \centering
    \includegraphics[width=8.4cm]{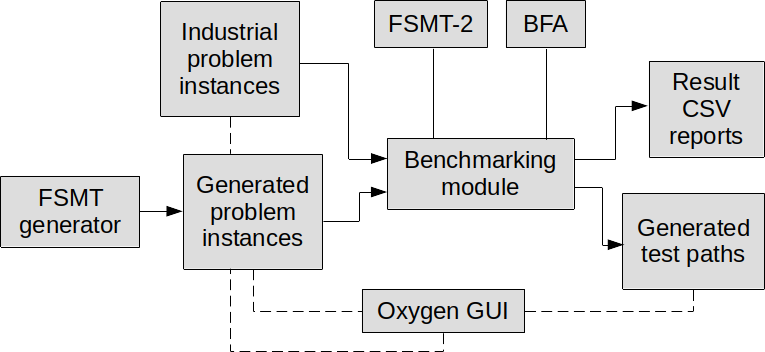}
    \caption{The infrastructure used for the experiments.}
    \label{fig:experiment_set_up}
\end{figure}

We used a benchmarking module that is part of the Oxygen platform. This module allows comparing individual algorithms that compute the test paths for a set of SUT models. The selected set of algorithms run for individual SUT models. The generated test paths are recorded in an Oxygen project. At the same time, the benchmarking module determines the defined properties of the generated test paths and can export them in a special CSV format to allow further analysis and processing of the experimental results. The whole experimental set-up is depicted in Figure \ref{fig:experiment_set_up}.

As explained before, in this experiment, we used two types of \textbf{problem instances}, anonymized and modified industrial FSMs from Skoda Auto and also artificially generated problem instances, generated by a special \textbf{FSMT generator} (details follow in Section \ref{sec:problem_instances}). These problem instances, together with the input parameters $minLength$, $maxLength$, and $testCoverage$, are an input to \textbf{benchmarking module}. Two algorithms, FSMT and NSR, are connected to this module. Then, all $P$ is generated for all problem instances in the input. The benchmarking module produces CSV reports with the properties of $P$ described in this section and enriches the Oxygen project files with the problem instances by these generated test paths.

\subsection{The used problem instances} \label{sec:problem_instances}

In the experiments, we used a mix of two types of problem instances: modifications of real industrial project state machine models and artificial SUT models generated by a special tool. In the presented results, we used six initial FSM-based models created by Skoda Auto test engineers. These models describe various parts of tested cars\footnote{Due to the confidentiality and Non-disclosure agreement, the types and brands of the tested cars were not mentioned in this paper.}. We further modified these models by removing the names of the states and transitions and slight modifications of each FSM to create four different problem instances. These modifications included adding cycles to an FSM, adding possible test starts and test ends, adding or removing a state, and adding and removing a transition. The result was 24 problem instances ($\boldsymbol{G}$) for a set of initial experiments. The selected properties of these instances are presented in Table \ref{tab:problem_instances_skoda}. For individual properties, minimal, maximal, average, and median values are given. 

In Table \ref{tab:problem_instances_skoda}, $cycles$ denotes the number of $\boldsymbol{G}$ cycles, $avg \: cycle \: length$ denotes the average length of these cycles. The $parallel \: edge \: groups$ denotes the number of groups of parallel edges present in $\boldsymbol{G}$, $parallel \: edges$ denotes the total number of parallel edges in $\boldsymbol{G}$, and $avg \: D+$ denotes the average node incoming degree. The $avg \: D-$ denotes the average node outgoing degree, and $avg \: D$ denotes the average node degree. By $|V_{ts} \cap V_{te}|$ we denote states in which a test path can both start and end.

\begin{table}%[h!]
\centering
\caption{Properties of problem instances created from industry project FSMs.}
%\begin{adjustbox}{max width=\textwidth}
\begin{tabular}{|c|c|c|c|c|}
\hline
\rowcolor{gray!25} \textbf{metric} & \textbf{min} & \textbf{max} & \textbf{average} & \textbf{median}\\
\hline

\cellcolor{gray!25}$|V|$ & 31 & 57 & 40.7 & 38\\ \hline
\cellcolor{gray!25}$|E|$ & 41 & 95 & 64.8 & 66.5\\ \hline
\cellcolor{gray!25}$cycles$ & 0 & 18 & 6.5 & 5.5\\ \hline
\cellcolor{gray!25}$avg \: cycle \: lenght$ & 0 & 22 & 4.1 & 5.2\\ \hline
\cellcolor{gray!25}$|V_{e}|$ & 1 & 21 & 8.1 & 4.5\\ \hline
\cellcolor{gray!25}$parallel \: edges$ & 0 & 18 & 4 & 1\\ \hline
\cellcolor{gray!25}$parallel \: edge \: groups$ & 0 & 9 & 2 & 0.5\\ \hline
\cellcolor{gray!25}$avg \: D+$ & 1 & 2 & 1.6 & 1.7\\ \hline
\cellcolor{gray!25}$avg \: D-$ & 1 & 2 & 1.6 & 1.7\\ \hline
\cellcolor{gray!25}$avg \: D$ & 2.1 & 4.1 & 3.2 & 3.4\\ \hline
\cellcolor{gray!25}$|V_{ts}|$ & 1 & 17 & 7.5 & 6.5\\ \hline
\cellcolor{gray!25}$|V_{te}|$ & 1 & 25 & 8.4 & 5\\ \hline
\cellcolor{gray!25}$|V_{ts} \cap V_{te}|$ & 0 & 6 & 1.5 & 1.5\\ \hline

\cellcolor{gray!25} SINGLE \textit{type defects} & 6 & 49 & 19.6\ & 18 \\ \hline

\cellcolor{gray!25} PAIR \textit{type defects} & 5 & 57 & 16.7 & 13 \\ \hline

\cellcolor{gray!25} $e_i \: to \: e_a \: distance$ & 0.6 & 2.8 & 1.6 & 1.6 \\ \hline

\end{tabular}
%\end{adjustbox}

\label{tab:problem_instances_skoda}
\end{table}

We generated an additional set of problem instances by ModelGen, a specialized module of the Oxygen platform. One of the functionality of this module is the generation of $\boldsymbol{G}$ problem instances by expected properties of the graph entered as input. These properties include: $|V|$, $|E|$, number of $\boldsymbol{G}$ cycles, $|V_{ts}|$,  $|V_{te}|$, $|V_{ts} \cap V_{te}|$, and $|V_{e}|$. For the experiments, we generated another 147 problem instances, varying by their properties as given in Table \ref{tab:generated_problem_instances}. The meaning of the metrics is the same as in Table \ref{tab:problem_instances_skoda}.

\begin{table}%[h!]
\centering
\caption{Properties of artificially generated problem instances.}
%\begin{adjustbox}{max width=\textwidth}
\begin{tabular}{|c|c|c|c|c|}
\hline
\rowcolor{gray!25} \textbf{metric} & \textbf{min} & \textbf{max} & \textbf{average} & \textbf{median}\\
\hline

\cellcolor{gray!25}$|V|$ & 15 & 23 & 17.7 & 15\\ \hline
\cellcolor{gray!25}$|E|$ & 23 & 35 & 31 & 35\\ \hline
\cellcolor{gray!25}$cycles$ & 2 & 3 & 2.5 & 2.5\\ \hline
\cellcolor{gray!25}$avg \: cycle \: lenght$ & 4 & 30.7 & 10.3 & 9\\ \hline
\cellcolor{gray!25}$|V_{e}|$ & 1 & 1 & 1 & 1\\ \hline
\cellcolor{gray!25}$parallel \: edges$ & 0 & 0 & 0 & 0\\ \hline
\cellcolor{gray!25}$parallel \: edge \: groups$ & 0 & 0 & 0 & 0\\ \hline
\cellcolor{gray!25}$avg \: D+$ & 1.5 & 2.3 & 1.8 & 1.5\\ \hline
\cellcolor{gray!25}$avg \: D-$ & 1.5 & 2.3 & 1.8 & 1.5\\ \hline
\cellcolor{gray!25}$avg \: D$ & 3 & 4.7 & 3.6 & 3.1\\ \hline
\cellcolor{gray!25}$|V_{ts}|$ & 1 & 2 & 1.5 & 1.5\\ \hline
\cellcolor{gray!25}$|V_{te}|$ & 1 & 2 & 1.5 & 1.5\\ \hline
\cellcolor{gray!25}$|V_{ts} \cap V_{te}|$ & 0 & 2 & 1 & 1\\ \hline

\cellcolor{gray!25}\color{blue}SINGLE \textit{type defects}\color{black} & \color{blue}2\color{black} & \color{blue}11\color{black}  & \color{blue}5.8\color{black} & \color{blue}5\color{black} \\ \hline

\cellcolor{gray!25}\color{blue}PAIR \textit{type defects}\color{black} & \color{blue}1\color{black} & \color{blue}10\color{black} & \color{blue}5.3\color{black} & \color{blue}5\color{black} \\ \hline

\cellcolor{gray!25}\color{blue}$e_i \: to \: e_a \: distance$ & \color{blue}1.2\color{black} & \color{blue}4\color{black} & \color{blue}2.3\color{black} & \color{blue}2.3\color{black} \\ \hline

\end{tabular}
%\end{adjustbox}

\label{tab:generated_problem_instances}
\end{table}

In total, we created 171 problem instances as presented in Table \ref{tab:all_problem_instances}. We considered this sample extensive enough to carry out the first experiments with the proposed strategy.

\subsection{Measurement of defect detection potential of test path sets}

An important question regarding the effectiveness of the generated $P$ is its potential to detect possible defects present in a SUT. This potential typically grows with the number of path combinations present in $P$, but the exact relation is difficult to identify. We added two types of fictional defects into experimental problem instances to evaluate $P$ produced by the presented FSMT strategy and baseline NSR.

A defect present in an SUT must be activated by a $p \in P$ to allow its detection by a tester or an automated test. Defect of a SINGLE type is defined at an $e \in E \in \boldsymbol{G}$, and we consider it to be activated when a $p \in P$ visits $e$.

PAIR type defect simulates data consistency defects in an SUT. It is defined as $(e_i,e_a), e_i,e_a \in E \in \boldsymbol{G}$, where there exists a path from $e_i$ to $e_a$. Transition $e_i$ causes simulated inconsistency of data stored in the SUT and transition $e_a$ causes its defective behavior. To activate the defect, $p \in P$ must visit $e_i$ and then visit $e_a$.

%($SINGLE \: type \: defects$ and $PAIR \: type \: defects$)

The numbers of SINGLE and PAIR type of artificial defects in experimental problem instances are given in Table \ref{tab:all_problem_instances}. for all problem instances, in Table \ref{tab:problem_instances_skoda} for problem instances generated from industry project FSMs and in Table \ref{tab:generated_problem_instances} for artificially generated problem instances. In Table \ref{tab:problem_instances_skoda}, \ref{tab:generated_problem_instances} and  \ref{tab:all_problem_instances}, $e_i \: to \: e_a \: distance$ denotes the number of edges between $e_i$ and $e_a$, averaged for all problem instances.

In the evaluation of $P$ properties, we further analyze the numbers of activated simulated defects, denoted as $\mathcal{A}_{S}$ for the SINGLE type and $\mathcal{A}_{P}$ for the PAIR type. Then we measure the average number of simulated defects activated by one test path step, denoted as $\mathcal{E}_{S} = \frac{SINGLE \: activated}{steps}$ for the SINGLE type and $\mathcal{E}_{P} = \frac{PAIR \: activated}{steps}$ for the PAIR type.

%\color{orange}

%Pri generovani vzdalenost 2-4

%Pocet type 1 a type 2 - 0.25x pocet hran

%\color{black}

\begin{table}%[h!]
\centering
\caption{Properties of all problem instances used in the experiments.}
%\begin{adjustbox}{max width=\textwidth}
\begin{tabular}{|c|c|c|c|c|}
\hline
\rowcolor{gray!25} \textbf{metric} & \textbf{min} & \textbf{max} & \textbf{average} & \textbf{median}\\
\hline

\cellcolor{gray!25}$|V|$ & 15 & 57 & 21 & 15\\ \hline
\cellcolor{gray!25}$|E|$ & 23 & 95 & 35.8 & 35\\ \hline
\cellcolor{gray!25}$cycles$ & 0 & 18 & 3.1 & 3\\ \hline
\cellcolor{gray!25}$avg \: cycle \: lenght$ & 0 & 30.7 & 9.5 & 8.7\\ \hline
\cellcolor{gray!25}$|V_{e}|$ & 1 & 21 & 2 & 1\\ \hline
\cellcolor{gray!25}$parallel \: edges$ & 0 & 18 & 0.6 & 0\\ \hline
\cellcolor{gray!25}$parallel \: edge \: groups$ & 0 & 9 & 0.3 & 0\\ \hline
\cellcolor{gray!25}$avg \: D+$ & 1 & 2.3 & 1.8 & 1.5\\ \hline
\cellcolor{gray!25}$avg \: D-$ & 1 & 2.3 & 1.8 & 1.5\\ \hline
\cellcolor{gray!25}$avg \: D$ & 2.1 & 4.7 & 3.5 & 3.1\\ \hline
\cellcolor{gray!25}$|V_{ts}|$ & 1 & 17 & 2.4 & 2\\ \hline
\cellcolor{gray!25}$|V_{te}|$ & 1 & 25 & 2.5 & 2\\ \hline
\cellcolor{gray!25}$|V_{ts} \cap V_{te}|$ & 0 & 6 & 1.1 & 1\\ \hline

\cellcolor{gray!25}\color{blue}SINGLE \textit{type defects}\color{black} & \color{blue}2\color{black} & \color{blue}49\color{black} & \color{blue}8\color{black} & \color{blue}6\color{black} \\ \hline

\cellcolor{gray!25}\color{blue}PAIR \textit{type defects}\color{black} & \color{blue}1\color{black} & \color{blue}57\color{black} & \color{blue}7.1\color{black} & \color{blue}6\color{black} \\ \hline

\cellcolor{gray!25}\color{blue}$e_i \: to \: e_a \: distance$ & \color{blue}0.6\color{black} & \color{blue}4\color{black} & \color{blue}2.2\color{black} & \color{blue}2.3\color{black} \\ \hline

\end{tabular}
%\end{adjustbox}
\label{tab:all_problem_instances}
\end{table}

%%%%%%%%%%% Bestoun --- 30 November

\subsection{Experiment results and discussion}

As explained in Section \ref{sec:test_coverage_criteria}, for certain $\boldsymbol{G}$ in combination with certain test set length ranges $minLenght$ to $maxLenght$, $P$ might not exist. This situation can be solved by changing $minLenght$ and $maxLenght$, or adding more possible $V_{ts}$ and $V_{te}$ to $\boldsymbol{G}$. However, this effect was present in the experiments and detail of its extent is given in Table \ref{tab:experiment_for_how_many_returned_P}. For defined test set length ranges (see Table \ref{tab:experiment_length_ranges}), out of total 171 problem instances, $P$ was returned for 138 up to 152 instances, depending on the strategy and test coverage criterion, denoted as $N_{all}$ in Table \ref{tab:experiment_for_how_many_returned_P}. More detail is given separately for industrial ($N_{industry}$) and artificial ($N_{artificial}$) problem instances.

\begin{table*}%[h!]
\centering
\caption{Number of test path sets found for individual length ranges, strategies and test coverage criteria. LR stands for Length Range.}
%\begin{adjustbox}{max width=\textwidth}
\begin{tabular}{|c|c|c|c|c|c|}
\hline
\rowcolor{gray!25} \textbf{LR set ID} & \textbf{Strategy} & \textbf{Test Coverage} &  $N_{all}$ & $N_{industry}$ & $N_{artificial}$ \\
\hline
\cellcolor{gray!25} \textbf{1} & FSMT & \textit{FSMT-level-1} & 143 & 119 & 24\\
\hline
\cellcolor{gray!25} \textbf{1} & FSMT & \textit{FSMT-level-2}  & 143 & 119 & 24\\
\hline
\cellcolor{gray!25} \textbf{1} & NSR & \textit{FSMT-level-1}  & 138 & 119 & 19\\
\hline
\cellcolor{gray!25} \textbf{1} & NSR & \textit{FSMT-level-2}  & 138 & 119 & 19\\
\hline
\cellcolor{gray!25} \textbf{2} & FSMT & \textit{FSMT-level-1} & 148 & 124 & 24\\
\hline
\cellcolor{gray!25} \textbf{2} & FSMT & \textit{FSMT-level-2}  & 148 & 124 & 24\\
\hline
\cellcolor{gray!25} \textbf{2} & NSR & \textit{FSMT-level-1}  & 143 & 124 & 19\\
\hline
\cellcolor{gray!25} \textbf{2} & NSR & \textit{FSMT-level-2}  & 143 & 124 & 19\\
\hline

\cellcolor{gray!25} \textbf{3} & FSMT & \textit{FSMT-level-1} & 152 & 128 & 24\\
\hline
\cellcolor{gray!25} \textbf{3} & FSMT & \textit{FSMT-level-2}  & 152 & 128 & 24\\
\hline
\cellcolor{gray!25} \textbf{3} & NSR & \textit{FSMT-level-1}  & 147 & 128 & 19\\
\hline
\cellcolor{gray!25} \textbf{3} & NSR & \textit{FSMT-level-2}  & 147 & 128 & 19\\
\hline
\cellcolor{gray!25} \textbf{4} & FSMT & \textit{FSMT-level-1} & 147 & 123 & 24\\
\hline
\cellcolor{gray!25} \textbf{4} & FSMT & \textit{FSMT-level-2}  & 147 & 123 & 24\\
\hline
\cellcolor{gray!25} \textbf{4} & NSR & \textit{FSMT-level-1}  & 142 & 123 & 19\\
\hline
\cellcolor{gray!25} \textbf{4} & NSR & \textit{FSMT-level-2}  & 142 & 123  & 19\\
\hline

\end{tabular}
%\end{adjustbox}
\label{tab:experiment_for_how_many_returned_P}
\end{table*}

Table \ref{tab:results} shows the experimental results for FSMT and NSR the problem instances summarized in Table \ref{tab:all_problem_instances} and expected test path length ranges as specified in Table \ref{tab:experiment_length_ranges}. In Table \ref{tab:results}, the average values for all results are given and $diff$ is a value for NSR divided by a value for FSMT.

\begin{table}%[h!]
\centering
\caption{Overall experimental results for FSMT and NSR (averages for all problem instances).}
%\begin{adjustbox}{max width=\textwidth}
\begin{tabular}{|c|c|c|c|c|c|c|c|c|c|}
\hline
\rowcolor{gray!25} \textbf{Strategy} & ~$len$~ & ~$|P|$~ & $avlen$ & $unique$ & ~$ut$~ & \color{blue}~$\mathcal{A}_{S}$~\color{black} & \color{blue}~$\mathcal{A}_{P}$~\color{black} & \color{blue}~$\mathcal{E}_{S}$~\color{black} & \color{blue}~$\mathcal{E}_{P}$~\color{black} \\
\hline

\rowcolor{gray!40} \multicolumn{10}{|c|}{\textbf{Length range set 1:}  $minLength=2$, $maxLength=4$} \\ \hline

\rowcolor{gray!10} \multicolumn{10}{|c|}{\textit{FSMT-level-1 Coverage}} \\ \hline

\cellcolor{gray!25}NSR & 9.5 & 3.0 & 3.2 & 6.8 & 1.4 & \color{blue} 3.55 \color{black} & \color{blue} 0.25 \color{black} & \color{blue} 0.29 \color{black} & \color{blue} 0.017 \color{black} \\ \hline
\cellcolor{gray!25}FSMT & 6.0 & 2.5 & 2.6 & 5.1 & 1.1 & \color{blue} 2.28 \color{black} & \color{blue} 0.04 \color{black} & \color{blue} 0.35 \color{black} & \color{blue} 0.009 \color{black} \\ \hline
\cellcolor{gray!25}$diff$ & 1.6 & 1.2 & 1.3 & 1.3 & 1.2 & \color{blue} 1.6 \color{black} & \color{blue} 6.0 \color{black} & \color{blue} 0.8 \color{black} & \color{blue} 2.0 \color{black}\\ \hline

\rowcolor{gray!10} \multicolumn{10}{|c|}{\textit{FSMT-level-2 Coverage}} \\ \hline

\cellcolor{gray!25}NSR & 19.4 & 5.8 & 3.3 & 10.9 & 1.6 & \color{blue} 6.15 \color{black} & \color{blue} 0.58 \color{black} & \color{blue} 0.25 \color{black} & \color{blue} 0.023 \color{black}\\ \hline
\cellcolor{gray!25}FSMT & 21.3 & 7.6 & 2.9 & 13.6 & 1.5 & \color{blue} 7.08 \color{black} & \color{blue} 0.56 \color{black} & \color{blue} 0.27 \color{black} & \color{blue} 0.018 \color{black}\\ \hline
\cellcolor{gray!25}$diff$ & 0.9 & 0.8 & 1.1 & 0.8 & 1.1 & \color{blue} 0.9 \color{black} & \color{blue} 1.0 \color{black} & \color{blue} 0.9 \color{black} & \color{blue} 1.3 \color{black}\\ \hline

\rowcolor{gray!40} \multicolumn{10}{|c|}{\textbf{Length range set 2:}  $minLength=2$, $maxLength=6$} \\ \hline

\rowcolor{gray!10} \multicolumn{10}{|c|}{\textit{FSMT-level-1 Coverage}} \\ \hline

\cellcolor{gray!25}NSR & 15.1 & 3.4 & 4.4 & 9.2 & 1.6 & \color{blue} 4.48 \color{black} & \color{blue} 0.65 \color{black} & \color{blue} 0.25 \color{black} & \color{blue} 0.036 \color{black}\\ \hline
\cellcolor{gray!25}FSMT & 7.2 & 2.7 & 2.8 & 6.0 & 1.2 & \color{blue} 2.65 \color{black} & \color{blue} 0.11 \color{black} & \color{blue} 0.35 \color{black} & \color{blue} 0.013 \color{black}\\ \hline
\cellcolor{gray!25}$diff$ & 2.1 & 1.3 & 1.6 & 1.5 & 1.4 & \color{blue} 1.7 \color{black} & \color{blue} 5.9 \color{black} & \color{blue} 0.7 \color{black} & \color{blue} 2.7 \color{black}\\ \hline

\rowcolor{gray!10} \multicolumn{10}{|c|}{\textit{FSMT-level-2 Coverage}} \\ \hline

\cellcolor{gray!25}NSR & 33.4 & 7.0 & 4.5 & 14.2 & 2.1 & \color{blue} 7.63 \color{black} & \color{blue} 1.40 \color{black} & \color{blue} 0.20 \color{black} & \color{blue} 0.035 \color{black}\\ \hline
\cellcolor{gray!25}FSMT & 32.0 & 9.7 & 3.6 & 17.3 & 1.8 & \color{blue} 8.67 \color{black} & \color{blue} 0.87 \color{black} & \color{blue} 0.23 \color{black} & \color{blue} 0.024 \color{black}\\ \hline
\cellcolor{gray!25}$diff$ & 1.04 & 0.7 & 1.3 & 0.8 & 1.2 & \color{blue} 0.9 \color{black} & \color{blue} 1.6 \color{black} & \color{blue} 0.9 \color{black} & \color{blue} 1.4 \color{black}\\ \hline

\rowcolor{gray!40} \multicolumn{10}{|c|}{\textbf{Length range set 3:}  $minLength=2$, $maxLength=8$} \\ \hline

\rowcolor{gray!10} \multicolumn{10}{|c|}{\textit{FSMT-level-1 Coverage}} \\ \hline

\cellcolor{gray!25}NSR & 20.6 & 3.6 & 5.5 & 10.7 & 1.9 & \color{blue} 5.15 \color{black} & \color{blue} 1.19 \color{black} & \color{blue} 0.23 \color{black} & \color{blue} 0.048 \color{black}\\ \hline
\cellcolor{gray!25}FSMT & 7.8 & 2.8 & 3.0 & 6.4 & 1.2 & \color{blue} 2.71 \color{black} & \color{blue} 0.11 \color{black} & \color{blue} 0.34 \color{black} & \color{blue} 0.014 \color{black}\\ \hline
\cellcolor{gray!25}$diff$ & 2.6 & 1.3 & 1.8 & 1.7 & 1.6 & \color{blue} 1.9 \color{black} & \color{blue} 11.3 \color{black} & \color{blue} 0.7 \color{black} & \color{blue} 3.5 \color{black}\\ \hline

\rowcolor{gray!10} \multicolumn{10}{|c|}{\textit{FSMT-level-2 Coverage}} \\ \hline

\cellcolor{gray!25}NSR & 45.8 & 7.5 & 5.8 & 16.1 & 2.6 & \color{blue} 8.15 \color{black} & \color{blue} 1.97 \color{black} & \color{blue} 0.17 \color{black} & \color{blue} 0.040 \color{black}\\ \hline
\cellcolor{gray!25}FSMT & 37.8 & 10.4 & 4.0 & 19.3 & 1.9 & \color{blue} 9.21 \color{black} & \color{blue} 1.11 \color{black} & \color{blue} 0.21 \color{black} & \color{blue} 0.029 \color{black}\\ \hline
\cellcolor{gray!25}$diff$ & 1.2 & 0.7 & 1.4 & 0.8 & 1.3 & \color{blue} 0.9 \color{black} & \color{blue} 1.8 \color{black} & \color{blue} 0.8 \color{black} & \color{blue} 1.4 \color{black}\\ \hline

\rowcolor{gray!40} \multicolumn{10}{|c|}{\textbf{Length range set 4:}  $minLength=4$, $maxLength=8$} \\ \hline

\rowcolor{gray!10} \multicolumn{10}{|c|}{\textit{FSMT-level-1 Coverage}} \\ \hline

\cellcolor{gray!25}NSR & 21.7 & 3.4 & 6.2 & 10.7 & 2.0 & \color{blue} 5.37 \color{black} & \color{blue} 1.31 \color{black} & \color{blue} 0.22 \color{black} & \color{blue} 0.052 \color{black}\\ \hline
\cellcolor{gray!25}FSMT & 10.9 & 2.5 & 4.5 & 8.0 & 1.3 & \color{blue} 3.53 \color{black} & \color{blue} 0.36 \color{black} & \color{blue} 0.30 \color{black} & \color{blue} 0.034 \color{black}\\ \hline
\cellcolor{gray!25}$diff$ & 2.0 & 1.4 & 1.4 & 1.3 & 1.5 & \color{blue} 1.5 \color{black} & \color{blue} 3.6 \color{black} & \color{blue} 0.7 \color{black} & \color{blue} 1.5 \color{black}\\ \hline

\rowcolor{gray!10} \multicolumn{10}{|c|}{\textit{FSMT-level-2 Coverage}} \\ \hline

\cellcolor{gray!25}NSR & 46.6 & 7.0 & 6.4 & 15.7 & 2.7 & \color{blue} 8.30 \color{black} & \color{blue} 2.10 \color{black} & \color{blue} 0.17 \color{black} & \color{blue} 0.042 \color{black}\\ \hline
\cellcolor{gray!25}FSMT & 39.2 & 8.2 & 5.1 & 18.2 & 2.0 & \color{blue} 9.12 \color{black} & \color{blue} 1.53 \color{black} & \color{blue} 0.20 \color{black} & \color{blue} 0.039 \color{black}\\ \hline
\cellcolor{gray!25}$diff$ & 1.2 & 0.9 & 1.3 & 0.9 & 1.3 & \color{blue} 0.9 \color{black} & \color{blue} 1.4 \color{black} & \color{blue} 0.8 \color{black} & \color{blue} 1.1 \color{black}\\ \hline

\end{tabular}
%\end{adjustbox}
\label{tab:results}
\end{table}

\begin{table}%[h!]
\centering
\caption{Overall experimental results for FSMT and NSR for industrial problem instances separately (averages for all problem instances).}
%\begin{adjustbox}{max width=\textwidth}
\begin{tabular}{|c|c|c|c|c|c|c|c|c|c|}
\hline
\rowcolor{gray!25} \textbf{Strategy} & ~$len$~ & ~$|P|$~ & $avlen$ & $unique$ & ~$ut$~ & \color{blue}~$\mathcal{A}_{S}$~\color{black} & \color{blue}~$\mathcal{A}_{P}$~\color{black} & \color{blue}~$\mathcal{E}_{S}$~\color{black} & \color{blue}~$\mathcal{E}_{P}$~\color{black} \\
\hline

\rowcolor{gray!40} \multicolumn{10}{|c|}{\textbf{Length range set 1:}  $minLength=2$, $maxLength=4$} \\ \hline

\rowcolor{gray!10} \multicolumn{10}{|c|}{\textit{FSMT-level-1 Coverage}} \\ \hline

\cellcolor{gray!25}NSR & 25.6 & 7.8 & 3.4 & 16.9 & 1.4 & \color{blue} 10.57 \color{black} & \color{blue} 0.91 \color{black} & \color{blue} 0.30 \color{black} & \color{blue} 0.024 \color{black} \\ \hline
\cellcolor{gray!25}FSMT & 18.0 & 7.7 & 2.3 & 14.0 & 1.2 & \color{blue} 5.83 \color{black} & \color{blue} 0.09 \color{black} & \color{blue} 0.36 \color{black} & \color{blue} 0.007 \color{black} \\ \hline
\cellcolor{gray!25}$diff$ & 1.4 & 1.0 & 1.5 & 1.2 & 1.2 & \color{blue} 1.8 \color{black} & \color{blue} 10.5 \color{black} & \color{blue} 0.8 \color{black} & \color{blue} 3.5 \color{black} \\ \hline

\rowcolor{gray!10} \multicolumn{10}{|c|}{\textit{FSMT-level-2 Coverage}} \\ \hline

\cellcolor{gray!25}NSR & 58.6 & 17.0 & 3.5 & 31.0 & 1.8 & \color{blue} 20.78 \color{black} & \color{blue} 2.09 \color{black} & \color{blue} 0.23 \color{black} & \color{blue} 0.036 \color{black} \\ \hline
\cellcolor{gray!25}FSMT & 79.1 & 29.8 & 2.6 & 50.9 & 1.5 & \color{blue} 26.91 \color{black} & \color{blue} 2.52 \color{black} & \color{blue} 0.27 \color{black} & \color{blue} 0.028 \color{black} \\ \hline
\cellcolor{gray!25}$diff$ & 0.7 & 0.6 & 1.3 & 0.6 & 1.1 & \color{blue} 0.8 \color{black} & \color{blue} 0.8 \color{black} & \color{blue} 0.8 \color{black} & \color{blue} 1.3 \color{black} \\ \hline

\rowcolor{gray!40} \multicolumn{10}{|c|}{\textbf{Length range set 2:}  $minLength=2$, $maxLength=6$} \\ \hline

\rowcolor{gray!10} \multicolumn{10}{|c|}{\textit{FSMT-level-1 Coverage}} \\ \hline

\cellcolor{gray!25}NSR & 40.9 & 8.5 & 4.8 & 22.6 & 1.7 & \color{blue} 12.35 \color{black} & \color{blue} 1.78 \color{black} & \color{blue} 0.25 \color{black} & \color{blue} 0.041 \color{black} \\ \hline
\cellcolor{gray!25}FSMT & 21.2 & 8.3 & 2.4 & 16.5 & 1.2 & \color{blue} 6.96 \color{black} & \color{blue} 0.30 \color{black} & \color{blue} 0.36 \color{black} & \color{blue} 0.015 \color{black} \\ \hline
\cellcolor{gray!25}$diff$ & 1.9 & 1.0 & 2.0 & 1.4 & 1.4 & \color{blue} 1.8 \color{black} & \color{blue} 5.9 \color{black} & \color{blue} 0.7 \color{black} & \color{blue} 2.8 \color{black} \\ \hline

\rowcolor{gray!10} \multicolumn{10}{|c|}{\textit{FSMT-level-2 Coverage}} \\ \hline

\cellcolor{gray!25}NSR & 92.4 & 18.2 & 4.9 & 36.1 & 2.4 & \color{blue} 24.48 \color{black} & \color{blue} 4.61 \color{black} & \color{blue} 0.17 \color{black} & \color{blue} 0.042 \color{black} \\ \hline
\cellcolor{gray!25}FSMT & 100.5 & 34.0 & 2.9 & 57.7 & 1.7 & \color{blue} 31.26 \color{black} & \color{blue} 3.13 \color{black} & \color{blue} 0.25 \color{black} & \color{blue} 0.029 \color{black} \\ \hline
\cellcolor{gray!25}$diff$ & 0.9 & 0.5 & 1.7 & 0.6 & 1.4 & \color{blue} 0.8 \color{black} & \color{blue} 1.5 \color{black} & \color{blue} 0.7 \color{black} & \color{blue} 1.5 \color{black} \\ \hline

\rowcolor{gray!40} \multicolumn{10}{|c|}{\textbf{Length range set 3:}  $minLength=2$, $maxLength=8$} \\ \hline

\rowcolor{gray!10} \multicolumn{10}{|c|}{\textit{FSMT-level-1 Coverage}} \\ \hline

\cellcolor{gray!25}NSR & 54.8 & 8.9 & 6.2 & 25.5 & 2.0 & \color{blue} 15.48 \color{black} & \color{blue} 3.61 \color{black} & \color{blue} 0.23 \color{black} & \color{blue} 0.055 \color{black} \\ \hline
\cellcolor{gray!25}FSMT & 21.8 & 8.4 & 2.5 & 16.6 & 1.2 & \color{blue} 7.00 \color{black} & \color{blue} 0.17 \color{black} & \color{blue} 0.36 \color{black} & \color{blue} 0.009 \color{black} \\ \hline
\cellcolor{gray!25}$diff$ & 2.5 & 1.1 & 2.5 & 1.5 & 1.6 & \color{blue} 2.2 \color{black} & \color{blue} 20.8 \color{black} & \color{blue} 0.6 \color{black} & \color{blue} 5.9 \color{black} \\ \hline

\rowcolor{gray!10} \multicolumn{10}{|c|}{\textit{FSMT-level-2 Coverage}} \\ \hline

\cellcolor{gray!25}NSR & 117.3 & 17.7 & 6.4 & 36.9 & 3.0 & \color{blue} 25.09 \color{black} & \color{blue} 5.78 \color{black} & \color{blue} 0.14 \color{black} & \color{blue} 0.041 \color{black} \\ \hline
\cellcolor{gray!25}FSMT & 106.7 & 34.8 & 3.0 & 59.2 & 1.8 & \color{blue} 32.12 \color{black} & \color{blue} 3.35 \color{black} & \color{blue} 0.24 \color{black} & \color{blue} 0.029 \color{black} \\ \hline
\cellcolor{gray!25}$diff$ & 1.1 & 0.5 & 2.1 & 0.6 & 1.7 & \color{blue} 0.8 \color{black} & \color{blue} 1.7 \color{black} & \color{blue} 0.6 \color{black} & \color{blue} 1.4 \color{black} \\ \hline

\rowcolor{gray!40} \multicolumn{10}{|c|}{\textbf{Length range set 4:}  $minLength=4$, $maxLength=8$} \\ \hline

\rowcolor{gray!10} \multicolumn{10}{|c|}{\textit{FSMT-level-1 Coverage}} \\ \hline

\cellcolor{gray!25}NSR & 54.6 & 8.0 & 6.6 & 24.3 & 2.1 & \color{blue} 15.09 \color{black} & \color{blue} 3.61 \color{black} & \color{blue} 0.22 \color{black} & \color{blue} 0.055 \color{black} \\ \hline
\cellcolor{gray!25}FSMT & 29.8 & 7.1 & 4.1 & 20.6 & 1.3 & \color{blue} 9.04 \color{black} & \color{blue} 0.87 \color{black} & \color{blue} 0.30 \color{black} & \color{blue} 0.037 \color{black} \\ \hline
\cellcolor{gray!25}$diff$ & 1.8 & 1.1 & 1.6 & 1.2 & 1.6 & \color{blue} 1.7 \color{black} & \color{blue} 4.2 \color{black} & \color{blue} 0.7 \color{black} & \color{blue} 1.5 \color{black} \\ \hline

\rowcolor{gray!10} \multicolumn{10}{|c|}{\textit{FSMT-level-2 Coverage}} \\ \hline

\cellcolor{gray!25}NSR & 116.5 & 16.5 & 6.8 & 35.5 & 3.2 & \color{blue} 24.57 \color{black} & \color{blue} 5.78 \color{black} & \color{blue} 0.14 \color{black} & \color{blue} 0.041 \color{black} \\ \hline
\cellcolor{gray!25}FSMT & 109.8 & 25.3 & 4.3 & 53.2 & 1.9 & \color{blue} 29.91 \color{black} & \color{blue} 4.26 \color{black} & \color{blue} 0.21 \color{black} & \color{blue} 0.042 \color{black} \\ \hline
\cellcolor{gray!25}$diff$ & 1.1 & 0.6 & 1.6 & 0.7 & 1.6 & \color{blue} 0.8 \color{black} & \color{blue} 1.4 \color{black} & \color{blue} 0.6 \color{black} & \color{blue} 1.0 \color{black} \\ \hline

\end{tabular}
%\end{adjustbox}
\label{tab:results_industrial}
\end{table}

\begin{table}%[h!]
\centering
\caption{Overall experimental results for FSMT and NSR for artificial problem instances separately (averages for all problem instances).}

%\begin{adjustbox}{max width=\textwidth}
\begin{tabular}{|c|c|c|c|c|c|c|c|c|c|}
\hline
\rowcolor{gray!25} \textbf{Strategy} & ~$len$~ & ~$|P|$~ & $avlen$ & $unique$ & ~$ut$~ & \color{blue}~$\mathcal{A}_{S}$~\color{black} & \color{blue}~$\mathcal{A}_{P}$~\color{black} & \color{blue}~$\mathcal{E}_{S}$~\color{black} & \color{blue}~$\mathcal{E}_{P}$~\color{black} \\
\hline

\rowcolor{gray!40} \multicolumn{10}{|c|}{\textbf{Length range set 1:}  $minLength=2$, $maxLength=4$} \\ \hline

\rowcolor{gray!10} \multicolumn{10}{|c|}{\textit{FSMT-level-1 Coverage}} \\ \hline

\cellcolor{gray!25}NSR & 6.8 & 2.2 & 3.2 & 5.1 & 1.3 & \color{blue} 2.19 \color{black} & \color{blue} 0.13 \color{black} & \color{blue} 0.28 \color{black} & \color{blue} 0.016 \color{black} \\ \hline
\cellcolor{gray!25}FSMT & 4.0 & 1.6 & 2.6 & 3.6 & 1.1 & \color{blue} 1.60 \color{black} & \color{blue} 0.03 \color{black} & \color{blue} 0.34 \color{black} & \color{blue} 0.009 \color{black} \\ \hline
\cellcolor{gray!25}$diff$ & 1.7 & 1.4 & 1.2 & 1.4 & 1.2 & \color{blue} 1.4 \color{black} & \color{blue} 3.8 \color{black} & \color{blue} 0.8 \color{black} & \color{blue} 1.7 \color{black} \\ \hline

\rowcolor{gray!10} \multicolumn{10}{|c|}{\textit{FSMT-level-2 Coverage}} \\ \hline

\cellcolor{gray!25}NSR & 12.8 & 3.9 & 3.3 & 7.5 & 1.6 & \color{blue} 3.32 \color{black} & \color{blue} 0.29 \color{black} & \color{blue} 0.25 \color{black} & \color{blue} 0.021 \color{black} \\ \hline
\cellcolor{gray!25}FSMT & 11.7 & 3.9 & 3.0 & 7.3 & 1.4 & \color{blue} 3.25 \color{black} & \color{blue} 0.18 \color{black} & \color{blue} 0.27 \color{black} & \color{blue} 0.016 \color{black} \\ \hline
\cellcolor{gray!25}$diff$ & 1.1 & 1.0 & 1.1 & 1.0 & 1.1 & \color{blue} 1.0 \color{black} & \color{blue} 1.6 \color{black} & \color{blue} 0.9 \color{black} & \color{blue} 1.3 \color{black} \\ \hline

\rowcolor{gray!40} \multicolumn{10}{|c|}{\textbf{Length range set 2:}  $minLength=2$, $maxLength=6$} \\ \hline

\rowcolor{gray!10} \multicolumn{10}{|c|}{\textit{FSMT-level-1 Coverage}} \\ \hline

\cellcolor{gray!25}NSR & 10.8 & 2.5 & 4.4 & 6.9 & 1.6 & \color{blue} 3.02 \color{black} & \color{blue} 0.44 \color{black} & \color{blue} 0.25 \color{black} & \color{blue} 0.035 \color{black} \\ \hline
\cellcolor{gray!25}FSMT & 4.9 & 1.7 & 2.9 & 4.3 & 1.1 & \color{blue} 1.85 \color{black} & \color{blue} 0.07 \color{black} & \color{blue} 0.35 \color{black} & \color{blue} 0.013 \color{black} \\ \hline
\cellcolor{gray!25}$diff$ & 2.2 & 1.4 & 1.5 & 1.6 & 1.4 & \color{blue} 1.6 \color{black} & \color{blue} 6.0 \color{black} & \color{blue} 0.7 \color{black} & \color{blue} 2.6 \color{black} \\ \hline

\rowcolor{gray!10} \multicolumn{10}{|c|}{\textit{FSMT-level-2 Coverage}} \\ \hline

\cellcolor{gray!25}NSR & 23.5 & 5.1 & 4.5 & 10.6 & 2.0 & \color{blue} 4.50 \color{black} & \color{blue} 0.81 \color{black} & \color{blue} 0.20 \color{black} & \color{blue} 0.033 \color{black} \\ \hline
\cellcolor{gray!25}FSMT & 20.6 & 5.6 & 3.7 & 10.6 & 1.8 & \color{blue} 4.48 \color{black} & \color{blue} 0.45 \color{black} & \color{blue} 0.22 \color{black} & \color{blue} 0.023 \color{black} \\ \hline
\cellcolor{gray!25}$diff$ & 1.1 & 0.9 & 1.2 & 1.0 & 1.1 & \color{blue} 1.0 \color{black} & \color{blue} 1.8 \color{black} & \color{blue} 0.9 \color{black} & \color{blue} 1.4 \color{black} \\ \hline

\rowcolor{gray!40} \multicolumn{10}{|c|}{\textbf{Length range set 3:}  $minLength=2$, $maxLength=8$} \\ \hline

\rowcolor{gray!10} \multicolumn{10}{|c|}{\textit{FSMT-level-1 Coverage}} \\ \hline

\cellcolor{gray!25}NSR & 14.9 & 2.7 & 5.4 & 8.2 & 1.8 & \color{blue} 3.30 \color{black} & \color{blue} 0.76 \color{black} & \color{blue} 0.22 \color{black} & \color{blue} 0.047 \color{black} \\ \hline
\cellcolor{gray!25}FSMT & 5.5 & 1.8 & 3.1 & 4.7 & 1.1 & \color{blue} 1.94 \color{black} & \color{blue} 0.09 \color{black} & \color{blue} 0.33 \color{black} & \color{blue} 0.015 \color{black} \\ \hline
\cellcolor{gray!25}$diff$ & 2.7 & 1.5 & 1.7 & 1.7 & 1.6 & \color{blue} 1.7 \color{black} & \color{blue} 8.1 \color{black} & \color{blue} 0.7 \color{black} & \color{blue} 3.2 \color{black} \\ \hline

\rowcolor{gray!10} \multicolumn{10}{|c|}{\textit{FSMT-level-2 Coverage}} \\ \hline

\cellcolor{gray!25}NSR & 33.9 & 5.8 & 5.7 & 12.6 & 2.5 & \color{blue} 5.10 \color{black} & \color{blue} 1.28 \color{black} & \color{blue} 0.17 \color{black} & \color{blue} 0.040 \color{black} \\ \hline
\cellcolor{gray!25}FSMT & 26.3 & 6.3 & 4.2 & 12.6 & 1.9 & \color{blue} 5.09 \color{black} & \color{blue} 0.71 \color{black} & \color{blue} 0.20 \color{black} & \color{blue} 0.029 \color{black} \\ \hline
\cellcolor{gray!25}$diff$ & 1.3 & 0.9 & 1.3 & 1.0 & 1.3 & \color{blue} 1.0 \color{black} & \color{blue} 1.8 \color{black} & \color{blue} 0.8 \color{black} & \color{blue} 1.4 \color{black} \\ \hline

\rowcolor{gray!40} \multicolumn{10}{|c|}{\textbf{Length range set 4:}  $minLength=4$, $maxLength=8$} \\ \hline

\rowcolor{gray!10} \multicolumn{10}{|c|}{\textit{FSMT-level-1 Coverage}} \\ \hline

\cellcolor{gray!25}NSR & 16.2 & 2.6 & 6.2 & 8.4 & 2.0 & \color{blue} 3.55 \color{black} & \color{blue} 0.88 \color{black} & \color{blue} 0.22 \color{black} & \color{blue} 0.052 \color{black} \\ \hline
\cellcolor{gray!25}FSMT & 7.7 & 1.7 & 4.6 & 5.9 & 1.3 & \color{blue} 2.50 \color{black} & \color{blue} 0.27 \color{black} & \color{blue} 0.30 \color{black} & \color{blue} 0.034 \color{black} \\ \hline
\cellcolor{gray!25}$diff$ & 2.1 & 1.5 & 1.4 & 1.4 & 1.5 & \color{blue} 1.4 \color{black} & \color{blue} 3.3 \color{black} & \color{blue} 0.7 \color{black} & \color{blue} 1.5 \color{black} \\ \hline

\rowcolor{gray!10} \multicolumn{10}{|c|}{\textit{FSMT-level-2 Coverage}} \\ \hline

\cellcolor{gray!25}NSR & 35.0 & 5.4 & 6.3 & 12.5 & 2.6 & \color{blue} 5.26 \color{black} & \color{blue} 1.41 \color{black} & \color{blue} 0.17 \color{black} & \color{blue} 0.043 \color{black} \\ \hline
\cellcolor{gray!25}FSMT & 27.4 & 5.3 & 5.2 & 12.4 & 2.1 & \color{blue} 5.24 \color{black} & \color{blue} 1.02 \color{black} & \color{blue} 0.20 \color{black} & \color{blue} 0.039 \color{black} \\ \hline
\cellcolor{gray!25}$diff$ & 1.3 & 1.0 & 1.2 & 1.0 & 1.3 & \color{blue} 1.0 \color{black} & \color{blue} 1.4 \color{black} & \color{blue} 0.8 \color{black} & \color{blue} 1.1 \color{black} \\ \hline

\end{tabular}
%\end{adjustbox}
\label{tab:results_artificial}
\end{table}

Starting with \textbf{FSMT-level-1 Coverage}, FSMT outperformed NSR in parameters $len$, $|P|$ and $avlen$ for the four test path length ranges examined. Taking into account \color{black} $len$, the total number of FSM transitions (test steps) in a test set, which is the parameter that gives the closest idea of the effort needed to execute the test paths, the difference between the strategies changed with the test path length range interval. For the test path length range set ID 1 (see Table \ref{tab:experiment_length_ranges}) where $maxLength-minLength=2$, the $diff$ was 1.6. For length range sets ID 2 and 4, where $maxLength-minLength=4$, differences were 2.1 and 2.0. For length range sets ID 3 where $maxLength-minLength=6$, the difference was the largest, 2.6. No such trend is obvious for $|P|$ in relation to the expected test path length difference. However, as expected, this trend is present for $avlen$ in the same way for $len$ (difference increasing from 1.3 to 1.8 with growing $maxLength-minLength$).

The test sets produced by NSR contain more unique FSM transitions, which is a consequence of the fact that these sets contain more transitions in general. Regarding $ut$, which measures the extent to how many FSM transitions have to be repeated to test one unique FSM transition, the results for FSMT are better than for NSR. The difference in $ut$ also increases with $maxLength-minLength$. However, not as obviously as in the case of $len$.

To give an overall summary, averaged by all test path length ranges, for \textit{FSMT-level-1 Coverage}, FSMT produced test paths with approximately one-half of the total steps than NSR and approximately by 25\% lower number of test paths in $P$.

Regarding the potential of test path sets to detect artificial defects inserted into SUT models, overall, longer test paths generated by NSR detected more SINGLE and PAIR type defects ($\mathcal{A}_{S}$ and $\mathcal{A}_{P}$ in Table \ref{tab:results}). This is a natural effect, and to evaluate the effectiveness of the set of test paths, the indicators $\mathcal{E}_{S}$ and $\mathcal{E}_{P}$ must be analyzed. Here, FSMT constantly outperformed NSR in the detection of SINGLE type defects ($\mathcal{E}_{S}$) for all length range sets, $diff$ ranging from 0.7 to 0.8 (for evaluation of artificial defects, smaller $diff$ means better result). This is a significant result - test path sets generated by FSMT detect approximately 20-30\% more defects per one test path step than NSR.

On the contrary, NSR outperforms FSMT in effectiveness in detecting PAIR type defects ($\mathcal{E}_{P}$) and $diff$ ranges from 1.5 to 3.5. However, this result has to be interpreted in the context of the number of detected PAIR type defects, which is very low compared to the SINGLE type. The results suggest that the state-machine-based testing technique with test coverage criteria as defined in this study is potentially ineffective for such a type of defect. Considering the results for baseline NSR in this aspect, the question is if a state-machine-based testing technique, in general, is effective in detecting PAIR type defects. However, the answer is beyond the scope of this study. For PAIR type defects, alternative techniques based on the life-cycle of data objects, e.g. the Data Cycle Test (DCyT) \cite{vroon2013tmap}, are available.

%%%%%%%%%%% Bestoun --- 30 November

For \textbf{FSMT-level-2 Coverage} (which subsumes \textit{FSMT-level-1 Coverage} criterion), \color{black} the results of FSMT are better than those of NSR. However, the difference is not so significant as in the case of \textit{FSMT-level-1 Coverage}. Regarding $len$, no significant differences are present for test path length range sets ID 1 and 2 having $maxLength \leq 6$. But the difference is 1.2 for sets ID 3 and 4 of the length range having $maxLength=8$. Here, for more complex test path generation problems, FSMT starts outperforming NSR.

Regarding the number of test paths $|P|$, FSMT produces a slightly higher number of test paths, on average 25\%. No clear trend is observed in relation to the testing path length range. Consequently, the average length of test paths ($avlen$) is, on average, 30\% lower for FSMT.

Taking into account the presence of unique FSM transitions in the test paths measured by $ut$, FSMT gives a slightly better result than NSR for $maxLength=6$, where the difference is 1.2, which further increases to 1.3 for $maxLength=8$.

To summarize, for \textit{FSMT-level-2 Coverage}, FSMT produced test paths having approximately 20\% fewer steps than the test paths produced by NSR for test path length ranges that have $maxLength=8$. On the contrary, no significant differences are observed for $maxLength \leq 6$. Regarding the overall number of these test paths, FSMT produced test sets with approximately 25\% more test paths than NSR. Lower $len$ practically implies lower testing costs, and at this point, this metric is much more significant than $|P|$. Hence, for one-half of examined cases ($maxLength=8$). We can conclude that FSMT outperformed NSR, and for the second half, there is no significant difference between the results of the algorithms. To this end, it is worth noticing that \textit{FSMT-level-2 Coverage} subsumes \textit{FSMT-level-1 Coverage} and is designed for more intense tests.

For \textit{FSMT-level-2 Coverage}, FSMT outperformed NSR in the detection of SINGLE type defects ($\mathcal{E}_{S}$) for all length range sets, $diff$ ranging from 0.8 to 0.9, practically meaning that $P$ generated by FSMT detect approximately 10-20\% more defects per one test path step than NSR. NSR outperformed FSMT in effectiveness to detect PAIR type defects ($\mathcal{E}_{P}$), $diff$ ranging from 1.1 to 1.4. It is noticeable that this difference is much smaller than in the case of \textit{FSMT-level-1 Coverage} criteria, but regarding the low number of PAIR type defects, the result for SINGLE type defect is much more significant.

The presented results showed good performance of the proposed FSMT strategy compared to the NSR strategy. The results show that a strategy such as NSR, based on the generation of all possible \textit{N-switch Coverage} test paths and their subsequent filtering, is not optimal to generate a test set satisfying \textit{FSMT-level-1} and \textit{FSMT-level-2} criteria. Our FSMT is needed to construct the test paths.

Particular data and differences for industrial and artificial problem instances separately are given in Tables \ref{tab:results_industrial} and \ref{tab:results}. However, the trends in the data are very similar to the overall results discussed in this section for the properties of the set of test paths, as well as their potential to detect artificial defects for both SINGLE and PAIR types.

\begin{figure*}%[hbt!]
    \centering
    \includegraphics[width=13cm]{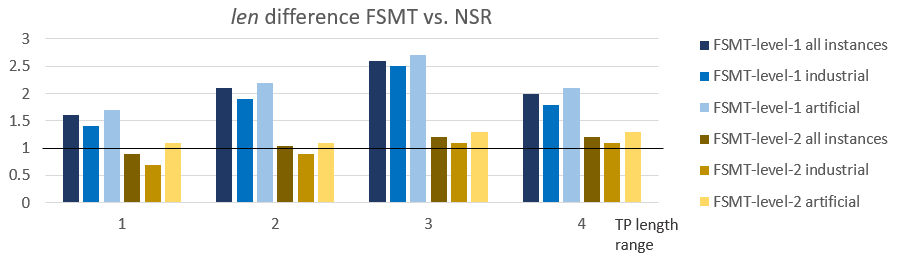}
    \caption{Difference in $len$ between FSMT and NSR for all problem instances together, then separately for industrial and artificial problem instances. \textit{FSMT-level-1} and \textit{FSMT-level-2} coverage criteria apply to both FSMT and NSR strategies.}
    \label{fig:graph_len_difference}
\end{figure*}

Overall summary of $len$, the main proxy for the testing costs is given in Figure \ref{fig:graph_len_difference}. The difference in $len$ for FSMT and NSR is shown separately for all the four expected test path (TP) length ranges (specified in Table \ref{tab:experiment_length_ranges}) and for all instances of problems together, followed by instances of industrial problems and instances generated artificially.

As we consider the $len$ as the main indicator used in the experiments, its relation to the length of general test cases and the input parameter $minLength$ shall be mentioned. As we explained in Section \ref{sec:introduction}, too short test cases are considered suboptimal by test engineers. However, what is "too short" might differ from project to project; hence, we give the engineer the liberty to determine the minimal length of the test paths by the $minLength$ parameter. This minimal length is part of the test coverage criteria that the generated set of test path $P$ must satisfy. During the process of $P$ generation, the proposed FSMT strategy tries to minimize the total length of these test paths ($len$). However, $P$ must satisfy defined test coverage criteria, so its test paths cannot be shorter than $minLength$ specified by the test engineer.

\begin{figure*}%[hbt!]
    \centering
    \includegraphics[width=12.8cm]{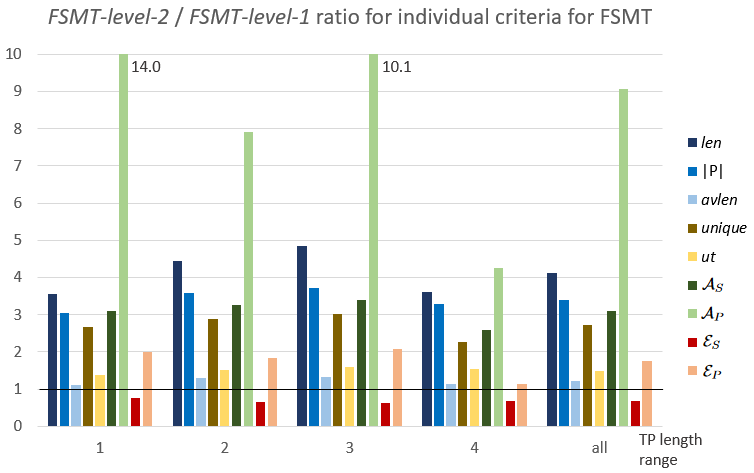}
    \caption{Comparison of averaged properties of all test path sets for \textit{FSMT-level-1} and \textit{FSMT-level-2} coverage criteria for the FSMT strategy.}
    \label{fig:coverage_criteria_difference_FSMT}
\end{figure*}

\begin{table}%[h!]
\centering
\caption{Comparison of test path set properties for \textit{FSMT-level-1} and \textit{FSMT-level-2 Coverage} levels for all problem instances}

%\begin{adjustbox}{max width=\textwidth}
\begin{tabular}{|c|c|c|c|c|c|c|c|c|c|}
\hline
\rowcolor{gray!25} \textbf{Strategy} & ~$len$~ & ~$|P|$~ & $avlen$ & $unique$ & ~$ut$~ & ~$\mathcal{A}_{S}$~ & ~$\mathcal{A}_{P}$~ & ~$\mathcal{E}_{S}$~  & ~$\mathcal{E}_{P}$~  \\
\hline

\rowcolor{gray!40} \multicolumn{10}{|c|}{\textbf{Length range set 1:}  $minLength=2$, $maxLength=4$} \\ \hline

\cellcolor{gray!25}NSR & 2.0 & 1.9 & 1.0 & 1.6 & 1.1 & 1.7 & 2.3 & 0.9 & 1.4 \\ \hline
\cellcolor{gray!25}FSMT & 3.6 & 3.0 & 1.1 & 2.7 & 1.4 & 3.1 & 14.0 & 0.8 & 2.0 \\ \hline

\rowcolor{gray!40} \multicolumn{10}{|c|}{\textbf{Length range set 2:}  $minLength=2$, $maxLength=6$} \\ \hline

\cellcolor{gray!25}NSR & 2.2 & 2.1 & 1.0 & 1.5 & 1.3 & 1.7 & 2.2 & 0.8  & 1.0 \\ \hline
\cellcolor{gray!25}FSMT & 4.4 & 3.6 & 1.3 & 2.9 & 1.5 & 3.3 & 7.9 & 0.7 & 1.8 \\ \hline

\rowcolor{gray!40} \multicolumn{10}{|c|}{\textbf{Length range set 3:}  $minLength=2$, $maxLength=8$} \\ \hline

\cellcolor{gray!25}NSR & 2.2 & 2.1 & 1.1 & 1.5 & 1.4 & 1.6 & 1.7 & 0.7 & 0.8 \\ \hline
\cellcolor{gray!25}FSMT & 4.8 & 3.7 & 1.3 & 3.0 & 1.6 & 3.4 & 10.1 & 0.6 & 2.1 \\ \hline

\rowcolor{gray!40} \multicolumn{10}{|c|}{\textbf{Length range set 4:}  $minLength=4$, $maxLength=8$} \\ \hline

\cellcolor{gray!25}NSR & 2.1 & 2.1 & 1.0 & 1.5 & 1.4 & 1.5 & 1.6 & 0.8 & 0.8 \\ \hline
\cellcolor{gray!25}FSMT & 3.6 & 3.3 & 1.1 & 2.3 & 1.5 & 2.6 & 4.3 & 0.7 & 1.1 \\ \hline

\rowcolor{gray!40} \multicolumn{10}{|c|}{\textbf{Average for length range set 1-4}} \\ \hline

\cellcolor{gray!25}NSR & 2.2 & 2.0 & 1.0 & 1.5 & 1.3 & 1.6 & 1.9 & 0.8 & 1.0 \\ \hline
\cellcolor{gray!25}FSMT & 4.1 & 3.4 & 1.2 & 2.7 & 1.5 & 3.1 & 9.1 & 0.7 & 1.8 \\ \hline

\end{tabular}

%\end{adjustbox}
\label{tab:comparison_of_coverage_criteria}
\end{table}

Table \ref{tab:comparison_of_coverage_criteria} compares averaged properties of $P$ generated by FSMT and NSR for the \textit{FSMT-level-1} and \textit{FSMT-level-2} coverage criteria for all problem instances. In Table \ref{tab:comparison_of_coverage_criteria}, ratio of averaged value of $P$ properties for all problem instances for \textit{FSMT-level-2} to this averaged value for \textit{FSMT-level-1} is presented. The last two lines of Table \ref{tab:comparison_of_coverage_criteria} present the average of these differences for all ranges of test path lengths.

For NSR, the difference between \textit{FSMT-level-2} and \textit{FSMT-level-1} is 2.2 on average in $len$ avergaed for all problem instances and 2 in $|P|$. For FSMT, this difference is 4.1 for $len$ and 3.4 for $|P|$. These differences have to be interpreted in the context of the average values $len$ and $|P|$ for the coverage criteria \textit{FSMT-level-2} and \textit{FSMT-level-1} separately (see Table \ref{tab:results}). As NSR produces $P$ with more test path steps and more test paths in general, and this difference is more significant for \textit{FSMT-level-1} test coverage, this effect is also reflected in the differences presented in Table \ref{tab:comparison_of_coverage_criteria}.

There is no significant difference in $avlen$ for NSR and a slight difference of 1.2 for FSMT. The difference between unique edges on the test paths ($unque$) is 1.5 for NSR and 2.7. for FSMT, which corresponds to the difference for $len$. Figure \ref{fig:coverage_criteria_difference_FSMT} presents the data analyzed for FSMT.

To summarize, test path sets that satisfy \textit{FSMT-level-2 Coverage} criterion that subsumes \textit{FSMT-level-1 Coverage} criterion generally consist of the approximately two times higher total number of steps in test paths for NSR and approximately four times for FSMT (although, the total number of steps in test path sets generated by FSMT does not exceed this number for NSR; see Table \ref{tab:results}).

Regarding the question, how much \textit{FSMT-level-1 Coverage} and \textit{FSMT-level-2 Coverage} criteria differ in the potential of test paths to detect defects, the differences in $\mathcal{E}_{S}$ and $\mathcal{E}_{P}$ in Table \ref{tab:comparison_of_coverage_criteria} must be analyzed. The results suggest that, on average,  test path sets that satisfy \textit{FSMT-level-1 Coverage} detect approximately 30\% more SINGLE type defects per one test path step for FSMT and 20\% more for NSR. However, this fact has to be interpreted in proper context; despite this result, test path sets satisfying \textit{FSMT-level-2 Coverage} detect more defects in total (see $\mathcal{A}_{S}$ in Table \ref{tab:comparison_of_coverage_criteria}). Regarding the low potential of both FSMT and NSR to detect PAIR type defects, we consider the difference for $\mathcal{E}_{P}$ to be insignificant.

\section{Threats to validity}\label{sec:threats_to_validity}

In this set of experiments, a few threats may cause bias in the results. The first threat is whether the NSR used in the experiments is the best to compare with FSMT objectively. The second issue is whether a set of 171 problem instances used in the experiments is extensive enough. In the experiments, we used a combination of industrial and artificially generated FSMs with a wide variety of sizes and other properties as well as four expected test path length ranges (see Table \ref{tab:experiment_length_ranges}). The related question is whether the examined problem instances are close enough to real-world examples. In the experiments, we used 24 problem instances created by an independent industrial team from FSMs models for various parts of Skoda Auto cars. Taking into account the trends observed for all problem instances (see Table \ref{tab:results}) and comparing them with the trends observed for these industrial problem instances (see Table \ref{tab:results_industrial}) and the generated problem instances (see Table \ref{tab:results_artificial}) separately, the results and trends are very similar. Therefore, no significant bias shall be caused by the choice of the SUT models used in the experiments. 

The last threat is related to the selection of the appropriate criteria for the comparison. In this study, we presented properties of test sets based on their size that are good proxies for estimating the required test effort, which is one of the key aspects in the real industrial testing process. 
In the comparison, we also use the number of two types of defects detected by a test path step. However, defects used in the experimental evaluation are artificial and randomly distributed in SUT models; this fact has to be taken into account when drawing conclusions from the results.

\section{Conclusion}\label{sec:conclusion}

In this study, we proposed an MBT technique to generate test paths for FSM in an implemented strategy. The new strategy allows us to concurrently express the possible start and end of test paths in an FSM and generate those that have a length in the given interval. The already published literature may address these requirements separately but not concurrently. The practical applicability of the proposed approach has already been verified through several real models from the car industry. We have compared the proposed FSMT with the best comparable alternative, NSR. We evaluated data from 1368 runs in total. We used a combination of 171 problem instances, two coverage criteria, and four test path length ranges. For all problem instances and all test path length ranges, FSMT clearly outperformed NSR for \textit{FSMT-level-1 Coverage} where it produced test paths with approximately only 50\% of total steps compared to the test paths produced by NSR. Furthermore, the number of total test paths in $P$ produced by FSMT was approximately 25\% lower than for NSR.

For \textit{FSMT-level-2 Coverage} the difference in the total number of steps in test paths was not significant for test length ranges that have $maxLength \leq 6$, but relevant for "longer" test path length ranges having $maxLength=8$, where FSMT produced test paths with approximately less 20\%  total steps than the test paths produced by NSR. As a trade-off, FSMT produced test sets with approximately 25\% more test paths than NSR.

As the total number of steps in test paths is the most important indicator that has a direct impact on testing costs, we can consider that FSMT outperforms NSR in all situations examined for \textit{FSMT-level-1 Coverage} and one-half of examined situations for \textit{FSMT-level-2 Coverage}, were in the second half, there was no significant difference.

Regarding the potential of test path sets to detect artificial defects inserted in a SUT model, FSMT generated test path sets detected approximately 20-30\% more SINGLE type defects per test path step than NSR for \textit{FSMT-level-1 Coverage} criterion and 10-20\% for \textit{FSMT-level-2 Coverage} criterion. The number of detected PAIR type defects was generally very low and suggested potential inefficiency of this version of the state-machine-based testing technique to detect them.

The results show good applicability of the proposed FSMT in situations when possible test path starts and ends in FSM needs to be reflected and, concurrently, the length of the test paths have to be in a defined range.

\section*{Acknowledgments}

The project is supported by CTU in Prague internal grant SGS20/177/OHK3/3T/13 “Algorithms and solutions for automated generation of test scenarios for software and IoT systems.” The authors acknowledge the support of the OP VVV funded project CZ.02.1.01/0.0/0.0/16\_019 /0000765 “Research Center for Informatics.” Bestoun S. Ahmed has been supported by the Knowledge Foundation of Sweden (KKS) through the Synergi Project AIDA - A Holistic AI-driven Networking and Processing Framework for Industrial IoT (Rek:20200067).

%Bibliography
\bibliographystyle{unsrt}  
\bibliography{manuscript.bib}

\end{document}